\def\aj{AJ}%
\def\apj{ApJ}%
\def\apjl{ApJ}%
\def\apjs{ApJS}%
\def\ao{Appl.~Opt.}%
\def\aap{A\&A}%
\def\mnras{MNRAS}%
\def\memras{MmRAS}%
\def\pasp{PASP}%
\documentclass[traditabstract]{aa} % for the abstract without structuration 
\usepackage{longtable}
\usepackage{graphicx}
\usepackage{latexsym}
\usepackage{amssymb}
\usepackage{lscape}
\usepackage{txfonts}
\begin{document}
%\VerbatimFootnotes
%{
   \title{Red-channel (6000-8000 \AA) ~nuclear spectra of 376 local galaxies
   \thanks{Tables 3 and 4 and Figure 4 are only available in electronic form
at the CDS via anonymous ftp to cdsarc.u-strasbg.fr (130.79.128.5)
or via http://cdsweb.u-strasbg.fr/cgi-bin/qcat?J/A+A/}}

   \author{Giuseppe Gavazzi \inst{1}, Guido Consolandi \inst{1}, Massimo Dotti  \inst{1}, Matteo Fossati  \inst{2,3,1}, Giulia Savorgnan \inst{4},  Roberto Gualandi \inst{5}, Ivan Bruni\inst{5}
          }
   \institute{
    Dipartimento di Fisica G. Occhialini, Universit\`a di Milano- Bicocca, Piazza della Scienza 3, I-20126 Milano, Italy\\
    \email{giuseppe.gavazzi@mib.infn.it, g.consolandi1@campus.unimib.it, massimo.dotti@mib.infn.it} 
    \and
    Universit{\"a}ts-Sternwarte M{\"u}nchen, Scheinerstrasse 1, D-81679 M{\"unchen}, Germany\\	     
    \email {mfossati@mpe.mpg.de}
    \and
    Max-Planck-Institut f{\"u}r Extraterrestrische Physik, Giessenbachstrasse, D-85748 Garching, Germany
    \and
    Centre for Astrophysics and Supercomputing, Swinburne University of Technology, Hawthorn, Victoria 3122, Australia\\
    \email {gsavorgn@astro.swin.edu.au}
    \and
    Osservatorio astronomico di Bologna, Via Ranzani 1, I-40127 Bologna, Italy \\
    \email{roberto.gualandi@oabo.inaf.it, ivan.bruni@oabo.inaf.it}
             }

    \date{Received - Accepted}

  \abstract
  {We obtained long-slit optical spectra of the nuclear regions of 
  376 galaxies in the local Universe using the 1.5m Cassini telescope of Bologna Observatory.
  Of these spectra, 164 were either never taken before by the Sloan Digital Sky Survey (SDSS), or 
  given by the Nasa Extragalactic Database (NED). With these new spectra, we  
  contribute investigating the occurrence of active galactic nuclei (AGNs).
  Nevertheless, we stress that the present sample is by no means
  complete, thus, it cannot be used to perform any demographic study.
  Following the method presented in Gavazzi et al (2011), we classify the nuclear spectra using a six bin scheme: 
  SEY (Seyfert), sAGN (strong AGN), and wAGN (weak AGN) represent active galactic nuclei of different levels of activity;
  HII accounts for star-forming nuclei; 
  RET (retired) and PAS (passive) refer to nuclei with poor or no star-formation activity.
  The spectral classification is performed using
  the ratio of $\lambda$ 6584 [NII] to $\rm H\alpha$ lines and the equivalent width (EW) of $\rm H\alpha$ 
  versus $[NII]/\rm H\alpha$ (WHAN diagnostic introduced by Cid Fernandes and collaborators) 
  after correcting $\rm H\alpha$ for underlying absorption.
  The obtained spectra are made available in machine readable format via the Strasbourg Astronomical Data Center (CDS) and NED.} 
   
     \keywords{Galaxies: active; Galaxies: nuclei; Galaxies: Seyfert}
%
%________________________________________________________________
\authorrunning{Gavazzi et al.}
\titlerunning{Nuclear spectroscopy of galaxies} 
\maketitle

\section{Introduction}
The advent of the Sloan Digital Sky Survey (SDSS, York et al. 2000)
revolutionized the course of astronomy at the turn of the millennium.
However some residual incompleteness remains in the SDSS  spectroscopic database,
especially at the bright luminosity end, due to shredding of the large galaxies
and fiber conflict (Blanton et al. 2005a,b,c).  Mitigating this problem
is a task that even 1.5m class telescopes can contribute to.  With this
idea, we decided to continue the spectroscopic project that began in
2005 at the Loiano Observatory (Gavazzi et al. 2011) with an aim at
searching for previously unknown optically selected active galactic nuclei (AGNs) in the local
Universe\footnote{The first year master students of G.G. are
  invited annualy to participate in some observing runs at the 1.5m Loiano
  telescope, which are kindly provided by the Observatory of Bologna (It).}. 
  In 2012-2013, we obtained 127 new nuclear spectra,
bringing the total number of galaxies independently observed to 376.
These are spectra taken with the red-channel of the spectrograph,
namely between 6000 and 8000 \AA.  Since nuclear spectra of about half  
these galaxies were neither obtained by SDSS nor distributed by NED, 
we provide our new observations to NED in order to make them publicly available in FITS (Flexible Image Transport Sistem) format.

The classification of AGNs based on optical nuclear spectra is routinely (eg, Decarli et al. 2007, Reines et al. 2013) 
performed using the BPT (Baldwin, Phillips \& Terlevich) diagnostic diagram 
(Baldwin et al. 1981), which requires the measurement of four spectral lines: $\rm H\beta$, [OIII], $\rm H\alpha$, 
and [NII].
General AGNs are disentangled from nuclear starbursts using 
the ratio [NII]/$\rm H\alpha$ (where $\rm H\alpha$ must be corrected for any underlying stellar absorption, 
as stressed by Ho et al. 1997), 
while strong AGNs (sAGN) can be separated from the weaker (weak AGNs or wAGN) LINERs  (Low-Ionization Nuclear Emission-Line Region) 
using the ratio [OIII]/$\rm H\beta$.
However a recent 
two-line diagnostic diagram named WHAN, which is based on the $\rm [NII]/H\alpha$ ratio combined with 
the strength of the $\rm H\alpha$ line was introduced by Cid Fernandes et al. (2010, 2011) 
to disentangle 
strong and weak AGNs, believed to be triggered by supermassive black holes from ``fake AGNs'',
dubbed as retired galaxies, whose ionization mechanism is probably provided by their old stellar population
(Trinchieri \& di Serego Alighieri 1991, Binette et al. 1994, 
Macchetto et al. 1996, Stasi{\'n}ska et al. 2008, Sarzi et al. 2010, Capetti \& Baldi 2011).
For a nucleus to be considered 
ionized by a central black hole, it is necessary that the equivalent width (hereafter EW) of $\rm H\alpha$ exceeds 3 \AA ~ 
(Cid Fernandes et al. 2010, 2011).
This quantitative threshold has been, however, questioned by Gavazzi et al. (2011), who
adopted $\rm EW H\alpha\geq$ 1.5 \AA.

Given the two-line WHAN diagnostic diagram, the red-channel spectra presented in this work can contribute to increasing 
the number of known AGNs, especially 
those associated with nearby bright galaxies, most affected by the residual incompleteness of the SDSS spectral database. 
Nevertheless we reiterate that the present sample is not complete by any means,
thus inadequate to perform any demographic study.

The outline of this paper is as follows.  
In Section \ref{sample} we describe the galaxy sample.
In Section \ref{The data} we illustrate the observations taken at the
Loiano observatory and the data reduction procedures. The nuclear
spectra are given and classified in Section \ref{results}.
In Section \ref{sec:broad} we outline the fitting method applied to broad-line systems.

\section{The sample}
\label{sample}

\begin{figure}
\begin{centering}
\includegraphics[width=9cm]{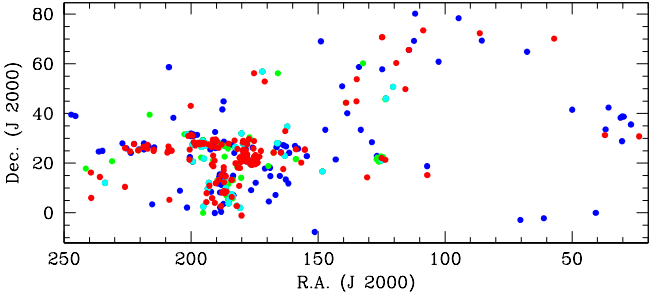}
\caption{Distribution in celestial coordinates of the 376 galaxies for which we obtained a nuclear spectrum. 
Color coding is as follows: available fiber spectra from SDSS (\#86, green);
available nuclear spectra from NED (\#162, blue); available spectra from both SDSS and NED (\#36, cyan); 
and newly obtained spectra from this work (\#164, red).
}
\label{radec}
\end{centering}
\end{figure}

\begin{figure}
\begin{centering}
\includegraphics[width=9cm]{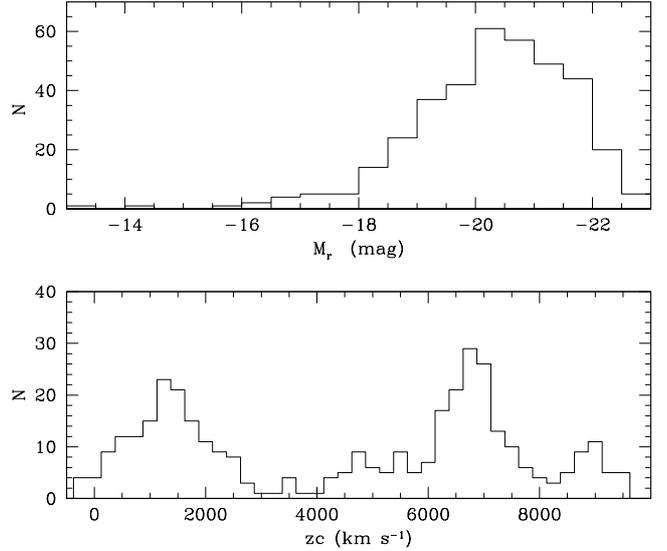}
\caption{Histograms of the absolute magnitudes ($r$-band and other bands mixed together)(top panel) 
and of the heliocentric recessional velocities (bottom panel) of the 376 studied galaxies.}
\label{histo}
\end{centering}
\end{figure}

This paper is based on a miscellaneous set of 376 northern galaxies visible in spring, which is not complete by any means, and therefore
not useful for performing any statistical study. 
They comprise of well-known bright local galaxies within $zc=10000 \rm ~km s^{-1}$, 
most of which have a nuclear spectrum already available from the literature,
mixed with fainter previously unobserved objects.
The latter were mainly selected from the Coma and Local 
superclusters (RA$\sim$ 180 deg, see Figure \ref{radec}) for not having a nuclear spectrum available from SDSS or for the 
being unavailable, in either ASCII or FITS form, 
in the nuclear optical spectra from NED\footnote{Except for 162, all 376 galaxies in our sample have a redshift from NED. This
does not necessarily mean that an optical nuclear spectrum is distributed by NED for all of them.}.
Figure \ref{histo} (top) illustrates the $r$-band absolute magnitude distribution of the selected galaxies, 
showing a dramatic lack of objects fainter than -20 mag with respect to to those predicted by the $r$-band luminosity function 
(e.g. Blanton et al. 2003).
Figure \ref{histo} (bottom) highlights the preferential membership of the selected targets to the Coma supercluster ($V\sim 7000 ~km s^{-1}$)
and the Virgo cluster and its surroundings ($V\sim 1000 ~km s^{-1}$).

\section{Observations and data reduction}
\label{The data}

We used the Bologna faint object spectrograph and camera
(BFOSC, Gualandi \& Merighi 2001) attached to the 152cm F/8 Cassini Telescope located
in Loiano, which belongs to the Observatory of Bologna, to obtain
optical spectra of the nuclei of 376 galaxies. The observations took place from 2005 to 2013 (see Table \ref{Table1}). 
The long-slit spectra were taken through a slit of 2 or 2.5 arcsec width (depending on the seeing conditions), 
with a intermediate-resolution red-channel grism (R $\sim$ 2200)
that covers the 6100 - 8200 \AA ~portion of the spectrum, which contains $\rm H\alpha$, [NII], and [SII] lines
(three galaxies were observed also using a  blue-channel grism covering H$\beta$ and [OIII], as
shown in Figure \ref{G7}).
The detector used by BFOSC is an EEV LN/1300-EB/1 CCD of 1300x1340 pixels, with 90\% quantum efficicency near 5500 \AA. 
Its spatial scale of 0.58 arcsec/pixel results in a field of view of $12.6'\times13'$.
The  dispersion of the red-channel grism is 8.8 nm/mm and results in spectra with 1.6 ~\AA/pix.
The instrumental broadening is typically $\sim$ 6 \AA ~full-width-half-maximum (FWHM), as checked on the 6300.3 \AA ~sky line.
We obtained exposures of 3-5 minutes, repeated typically three-six times per run (to remove the cosmic rays), 
but several galaxies were re-observed in more than one run (see Table \ref{data2}). The seeing at Loiano is typically 1.5 - 2.5 arcsec.
The slit was mostly set in the E-W direction, except when it was positioned along the galaxy major axis
or along the direction connecting two adjacent objects to accomodate both objects in one exposure. 
The wavelength was calibrated using
frequent exposures of a He-Ar hollow-cathode lamp. We used several sky lines to check a posteriori the wavelength calibration.
The spectrograph response was obtained by daily exposures of the star Feige-34.
\begin{table}[ht!]
\caption{Log of the observations at Loiano. Two new-moon periods of four nights were allocated to the present 
project per year. We list information for the useful 
nights. In total, we obtained 422 spectra of 376 independent galaxies. That is, several objects were repeatedly observed during different runs. 
In these cases, Table \ref{data1} and Figure \ref{spectra} describe the combined spectra (labeled L00).}
\centering
{\footnotesize \begin{tabular}{llll}
\hline
\hline
Year  & Feb           & Mar           &  Nspec \\
2005  &   9,10,11     &   9,11,12     &  45  \\      
2006  &   26          &   23          &  24  \\  
2007  &   13,15,16    &   17,18       &  29  \\  
2008  &   5           &   3           &  10  \\
2009  &   18,19,21    &   26,27       &  54  \\
2010  &   8           &    -          &  10  \\
2011  &   7,8,9,10    &   6,7,8       &  133 \\
2012  &   23,24       &   14,15,16,17 &  74  \\
2013  &   6,8,9       &   12,15       &  53  \\
\hline
\end{tabular}
}
\label{Table1}
\end{table}

  The spectra were reduced using the IRAF-STSDAS\footnote{IRAF is the Image Analysis and Reduction Facility made
  available to the astronomical community by the National Optical
  Astronomy Observatories, which are operated by AURA, Inc., under
  contract with the U.S. National Science Foundation. STSDAS is
  distributed by the Space Telescope Science Institute, which is
  operated by the Association of Universities for Research in
  Astronomy (AURA), Inc., under NASA contract NAS 5--26555.}
packages, following reduction procedures identical to Gavazzi et al. (2011)
that are not repeated here. 
In summary, the flux and wavelength-calibrated 1-D spectra were extracted in apertures of 5.8 arcsec,
normalized to the intensity of the continuum under $\rm
H\alpha$, transformed to rest-frame velocity using the redshift
provided by NED, and measured to obtain the intensity and EW of the $\rm H\alpha$ and [NII] lines.  

The presence of a strong
telluric absorption feature near 6866 \AA ~hampers the detection of
[SII] lines at redshift near $zc=6000-7000~\rm km~s^{-1}$. Therefore, these lines are
not analyzed\footnote{The same telluric feature would overlap with H$\alpha$
near $zc=13900~\rm km~s^{-1}$, which is larger than the largest redshift in our sample.}.
In the presence of broad permitted lines, the measurements of the EW and
FWHM of the broad and narrow emission lines were obtained using the
fitting procedure discussed in Section \ref{sec:broad}.

\section{Spectral classification}
\label{results}
The classification of the nuclear activity based on optical spectra
was performed following the methods of Gavazzi et al. (2011) 
based on the WHAN diagram (Cid Fernandes et al. 2010, 2011). 
In short, the WHAN classification scheme is based on 
the strength of the
$\rm H\alpha_{corr}$ line ($\rm H\alpha$, corrected for underlying stellar absorption, Ho et
al. 1997) and the ratio between the flux of [NII] and $\rm
H\alpha_{corr}$.
As in Gavazzi et al. (2011), we adopt a mean underlying stellar absorption at $\rm H\alpha$
of 1.3 $\AA$, irrespective of luminosity.  
%Thus we adopt 1.3 $\AA$ for the underlying
%continuum absorption at $\rm H\alpha$, i.e. $\rm EW H\alpha_{corr} =EW H\alpha - 1.3 ~\AA$
\footnote{Because $\rm H\alpha$ and [NII] are close one another
and since we use fluxes normalized to the continuum under $\rm H\alpha$, 
the same correction for stellar absorption is applied to the fluxes as to the EWs.}
 %\footnote{For star-forming galaxies, the direct
 %determination of absorption is hampered by the presence of $\rm
 %H\alpha$ and [NII] in emission.  However, for bright spirals ($\rm
 %Log (L_i/L_{\odot})>10$, often dominated by bulges) the presence of
 %underlying absorption at $\rm H\alpha$ is sometimes detected
 %(although difficult to quantify) in the SDSS spectra.  In contrast,
 %for faint late-types ($\rm Log (L_i/L_{\odot})\leq 10$) the
 %absorption component at $\rm H\alpha$ is generally not detected,
 %being overtaken by the emission component, which amounts on average
 %to 30 \AA~ (and exceeds 5 \AA~for 93\% of them), making the 1.3
 %~\AA~ correction negligible.}. \\
\begin{figure*}
\begin{centering}
\includegraphics[width=13cm]{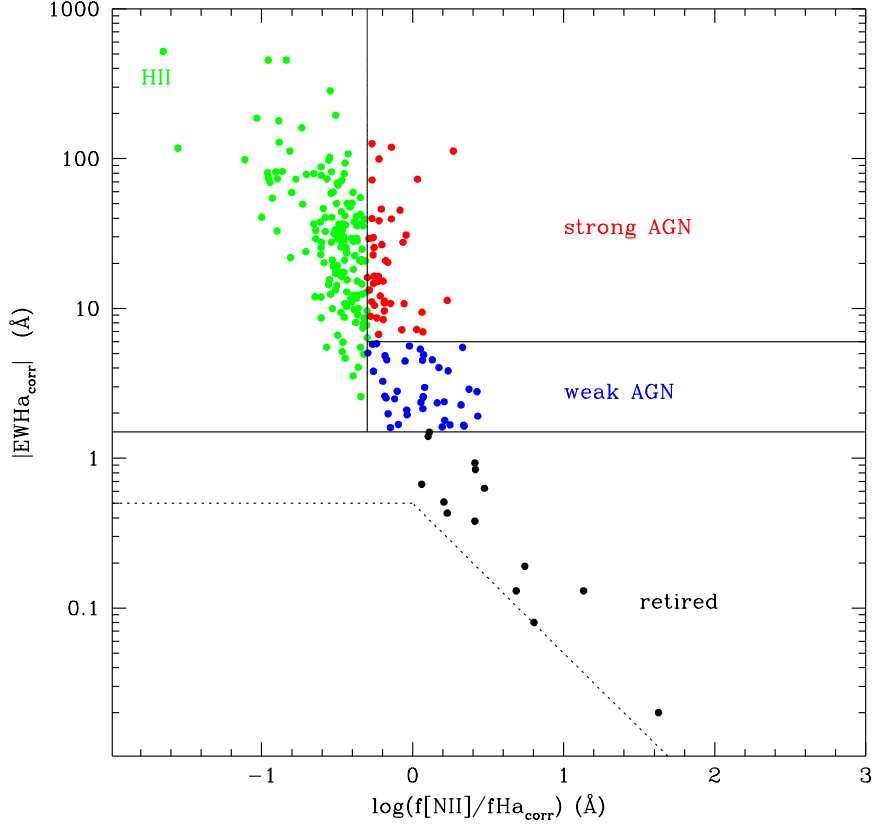}
\caption{The WHAN diagnostic diagram for nuclei whose $\rm H\alpha_{corr}$ line is detected in emission. 
  The equivalent width of $\rm H\alpha$ is corrected by 1.3 \AA~ for underlying absorption. Passive
  nuclei never appear in the figure, because they have either H$\rm \alpha_{corr}$ in
  absorption or undetected [NII].  The 4 SEY1 objects identified by their broad-line systems are not plotted.}
\label{whan}
\end{centering}
\end{figure*}

Figure \ref{whan} shows the WHAN diagram with the six adopted spectral
classification thresholds. These thresholds (whose detailed description can be obtained from Gavazzi et al. 2011) allow
to separate HII region-like nuclei (marked HII in Table \ref{data2}) from strong AGNs (sAGN), weak AGNs or LINERs (wAGN),
and fake AGNs (dubbed as "retired nuclei" or RET
whose ionization mechanism is provided by their old stellar population).
Passive nuclei (PAS) display no star formation activity. Their spectra show either
H$\rm \alpha_{corr}$ in absorption or undetected [NII].
Four Seyfert 1 galaxies are classified by visual inspection of the
individual spectra for the presence of broad permitted lines (see Section \ref{sec:broad}).

Table \ref{data1} gives the parameters of the 376 independently observed galaxies as follows:\\
Column 1: serial number;\\
Column 2: Galaxy name. The first three digits of the label give the year of observation. 
L00 marks those spectra that are a median of two or more spectra obtained in different runs.
Then the catalog name and serial number: Arecibo galaxy catalog (AGC) (Haynes et al.2011), SDSS (DR7, Abazajian et al.2009); 
catalog of galaxies and clusters of galaxies (CGCG) (Zwicky et al. 1968), Uppsala galaxy catalog (UGC) (Nilson 1973), 
new galaxy catalog (NGC) (Dreyer 1888) 
and index catalog (IC) (Dreyer 1908), Virgo cluster catalog (VCC) (Binggeli et al.1985), Markarian (MRK) (Markarian 1967), Herschel Reference Survey  
(HRS) (Boselli et al 2010) designations;\\
%Column 2 and 3: J2000 celestial coordinates (degrees);\\
Column 3 and 4: J2000 celestial coordinates;\\
Column 5: heliocentric redshift, as listed by NED;\\
Column 6: redshift independent distance in Mpc, as listed by NED. When this is unavailable, the 
galactocentric Hubble-flow distance is listed;\\
Column 7: $r$-band (AB) magnitude from SDSS when available. Otherwise, the total (Vega) magnitude
in the Johnson R, V or B band (as specified in parenthesis) from NED is used;\\
Column 8: a cross indicates the availability of a nuclear spectrum in NED;\\
Column 9: a cross indicates the availability of a fiber nuclear spectrum in SDSS.  

Table \ref{data2} summarizes the classification of spectra  as follows:\\
Column 1: serial number;\\
Column 2: Galaxy name;\\
Column 3: number of individual exposures;\\
Column 4: duration of the individual exposures;\\
Column 5: slit aperture in arcsec;\\
Column 6: slit orientation (counterclockwise from N) in the various runs (yy);\\
Column 7: measured EW  of $\lambda$ 6548.1 [NII] line (dubbed [NII]1)(negative EW values represent emission);\\
Column 8: measured EW of the (narrow) H$\alpha$ line;\\
Column 9: measured EW  of $\lambda$ 6583.6 [NII] line (dubbed [NII]2);\\
Column 10: R.M.S noise of the individual spectra determined between 6400 and 6500 \AA;\\
Column 11: six bins nuclear activity classification.

We note that the EW and FWHM of the narrow H$\alpha$ and of the two [NII]
lines in Seyfert 1 galaxies have been computed applying the procedure
described in Sect.~\ref{sec:broad} to consider the
underlying broad H$\alpha$ component.  Adding all 160 spectra of
HII-like nuclei, we obtain a template spectrum with a high
signal-to-noise ratio (2370 at H$\alpha$).  A ratio of
$[NII]_{6583}/[NII]_{6548}=3.11$ is obtained, which is consistent with the theoretical value
of 3.0 obtained by Osterbrock \& Ferland (2006).

%For high (100) signal-to-noise spectra le FHWHM of [NII2] and Halpha are correlated 
%and provide a measurement of the velocity dispersion inside the aperture.

\section{Broad line measurements}
\label{sec:broad}

Four (L09 UGC-1935, L12 MRK-0079, L11 NGC-3758E, and L13 MRK-0841) of
the 376 galaxies show the presence of a broad H$\alpha$ line. To
properly estimate the EW and FWHM of these broad lines and of the narrow H$\alpha$ and [NII] lines, both the broad and
narrow components have to be fit at the same time. We perform a very
simple minimum $\chi$-square fit, assuming that the underlying continuum
follows a power-law profile and that every single broad and narrow
line is well described by a single Gaussian profile.  As a disclaimer,
we note that the FWHM of the broad H$\alpha$ lines can be slightly
affected by a poor treatment of the underlying continuum, which does
not include any specific galactic feature and where the very low flux
tails of the lines are not fitted as well as possible by assuming two
broad Gaussian, a Lorentzian, or a Voigt profile. The EW is almost
unaffected by these small approximations. A high precision measurement
of the FWHM of the broad H$\alpha$ is beyond the scope of this analysis.

The fitting procedure has a total of nine free parameters: two describe
the power-law continuum, four are related to the narrow [NII] and H$\alpha$
lines, and three describe the broad H$\alpha$ line. More
specifically, these include\\ $\bullet$ the normalization and exponent of the power
low continuum;\\ $\bullet$ the three normalizations of the narrow
H$\alpha$ and [NII] lines and their FWHM, which are assumed to be the same for
every narrow line;\\ $\bullet$ the normalization, FWHM, and peak
frequency of the broad H$\alpha$ line. We let the peak frequency vary to
best fit the line profile, although no significant shifts have been
found.\\
\begin{table}[h]
\begin{center}
\caption{Broad H$\alpha$ line measurements.}
\begin{tabular}{lcc}
\hline
\hline
Obj		    & EWH$\alpha$     & FWHM   \\
        	    & \AA     & $\rm km~ s^{-1}$ \\
\hline
 L09 UGC-1935	    & -178.2  & 3480  \\
 L12 MRK-0079	    & -297.7  & 3706  \\
 L11 NGC-3758E      & -128.0  & 3426   \\
 L13 MRK-0841	    & -408.3  & 5042   \\
\hline
\hline
\end{tabular}
\label{Table5}
\end{center}
\end{table}
The numerical values of the FWHM and EW obtained for the  
broad lines are reported in Table~\ref{Table5}. For these four Seyfert 1 galaxies, the full fits
with the fits of only the broad component are superimposed to
the original spectra in Figure~\ref{broad}.

\begin{acknowledgements}
G. Gavazzi thanks his first year master students for their assistence
during many observing nights at Loiano.  
We thank Giorgio Calderone for discussions on the measurements of the
broad-line systems.   We are grateful
to Paolo Franzetti and Alessandro Donati for their contribution to
GoldMine, the Galaxy On Line Database (Gavazzi et al. 2003) extensively used in this work
(http://goldmine.mib.infn.it).
We thank Barry Madore for useful comments on the manuscript.
We also thank the coordinator of the TAC,
Valentina Zitelli for the generous time allocation.
This research has made use of the NASA/IPAC Extragalactic Database (NED) 
which is operated by the Jet Propulsion Laboratory, California Institute of Technology, 
under contract with the National Aeronautics and Space Administration. 
%We acknowledge the constructive criticism from an anonymous referee.\\
The present study made extensive use of the DR7 of SDSS. 
Funding for the Sloan Digital Sky Survey (SDSS) and SDSS-II has been provided by the 
Alfred P. Sloan Foundation, the Participating Institutions, the National Science Foundation, 
the U.S. Department of Energy, the National Aeronautics and Space Administration, 
the Japanese Monbukagakusho, and 
the Max Planck Society, and the Higher Education Funding Council for England. 
The SDSS Web site is http://www.sdss.org/.
The SDSS is managed by the Astrophysical Research Consortium (ARC) for the Participating Institutions. 
The Participating Institutions are the American Museum of Natural History, Astrophysical Institute Potsdam, 
University of Basel, University of Cambridge, Case Western Reserve University, The University of Chicago, 
Drexel University, Fermilab, the Institute for Advanced Study, the Japan Participation Group, 
The Johns Hopkins University, the Joint Institute for Nuclear Astrophysics, the Kavli Institute for 
Particle Astrophysics and Cosmology, the Korean Scientist Group, the Chinese Academy of Sciences (LAMOST), 
Los Alamos National Laboratory, the Max-Planck-Institute for Astronomy (MPIA), the Max-Planck-Institute 
for Astrophysics (MPA), New Mexico State University, Ohio State University, University of Pittsburgh, 
University of Portsmouth, Princeton University, the United States Naval Observatory, and the University 
of Washington.
% G. Gavazzi acknowledges financial support from Italian MIUR PRIN contract 200854ECE5.
\end{acknowledgements}

 \clearpage
 \begin{onecolumn}
 \tiny

 \begin{longtable}{clccccccc}
 \caption{{\bf General parameters for a representative sample of 20 galaxies. L00 means that spectra taken in more than one year were combined.
 The full set ot 376 measurements is available in electronic form at the CDS.}}\\
 \hline
 \noalign{\smallskip}
     ID&  Obj		      &  RA	     & Dec	   &   $z$   &   Dist	  &   $r$	&    NED    & SDSS	\\				%     rag	 deg	     HF
       &		      &  hh:mm:ss,ss & dd:pp:ss.s  &	      &  Mpc	  &    mag	&	    &		\\				%
   (1) &  (2)		      &  (3)	     & (4)	   &   (5)    &   (6)	  &   (7)	&    (8)    & (9)	\\				%
 \hline 																		%
 \endfirsthead
 \hline
 \noalign{\smallskip}	 ID&  Obj		      &  RA	     & Dec	   &   $z$   &   Dist	  &   $r$	&    NED    & SDSS	\\				%     rag deg	      HF
       &		      &  hh:mm:ss,ss & dd:pp:ss.s  &	      &  Mpc	  &    mag	&	    &		\\				%
   (1) &  (2)		      &  (3)	     & (4)	   &   (5)    &   (6)	  &   (7)	&    (8)    & (9)	\\				%
 \hline 																	%
 \endhead
 %  & L13-NGC-0205	  &  00:40:22.08 & +41:41:07.1 &-0.000804 &    0.79   &  10.69      &	  x	&  -	    \\  			    %	 10.0920000   41.685306    -
 %  & L13-M-32  	  &  00:42:41.83 & +40:51:55.0 &-0.000667 &    0.77   &   8.29      &	  x	&  -	    \\  			    %	 10.6743000   40.865286    -
 %  & L13-M-31  	  &  00:42:44.35 & +41:16:08.6 &-0.001001 &    0.79   &   3.44  (V) &	  x	&  -	    \\  			    %	 10.6847930   41.269065    -
     1 & L11-M33-HII	      &  01:34:33.19 & +30:47:00.4 &-0.000597 &    0.88   &    -	&	    &		\\				%    23.6382800   30.783460    -
     2 & L05-CGCG-522-004     &  01:47:16.15 & +35:33:47.9 & 0.015591 &    78.3   &  12.32  (V) &     x     &		\\				%    26.8172920   35.563306    -
     3 & L05-CGCG-522-093     &  01:58:35.20 & +38:43:06.8 & 0.016878 &    71.2   &  15.21  (V) &     x     &		\\				%    29.6466670   38.718556    HF
     4 & L05-CGCG-522-104     &  02:00:59.71 & +38:47:04.7 & 0.018880 &    79.4   &  15.15  (V) &     x     &		\\				%    30.2488030   38.784637    HF
     5 & L11-NGC-0784	      &  02:01:16.93 & +28:50:14.1 & 0.000660 &    4.62   &  12.16    - &     x     &		\\				%    30.3205420   28.837250    -
     6 & L05-CGCG-522-106     &  02:03:44.83 & +38:15:31.4 & 0.019227 &    62.3   &  13.09  (V) &     x     &		\\				%    30.9367920   38.258722    -
     7 & L09-NGC-0891	      &  02:22:33.41 & +42:20:56.9 & 0.001761 &    10.2   &   7.86  (R) &     x     &		\\				%    35.6392240   42.349146    -
     8 & L00-NGC-0925	      &  02:27:16.88 & +33:34:45.0 & 0.001845 &    8.57   &   9.55  (R) &     x     &		\\				%    36.8203330   33.579166    -
     9 & L09-UGC-1935	      &  02:28:14.47 & +31:18:41.9 & 0.016652 &    52.3   &  14.64	&	    &		\\				%    37.0603170   31.311658    -
    10 & L00-NGC-1068	      &  02:42:40.71 & -00:00:47.8 & 0.003793 &    13.5   &  10.62	&     x     &		\\				%    40.6696290   -0.013280    -
    11 & L00-NGC-1275	      &  03:19:48.16 & +41:30:42.1 & 0.017559 &    68.2   &  11.61  (R) &     x     &		\\				%    49.9506670   41.511696    -
    12 & L09-UGC-2855	      &  03:48:20.73 & +70:07:58.4 & 0.004003 &    17.1   &  11.11  (R) &	    &		\\				%    57.0863750   70.132889    -
    13 & L11-NGC-1507	      &  04:04:27.21 & -02:11:18.9 & 0.002879 &    11.1   &  11.90  (R) &     x     &		\\				%    61.1133640   -2.188577    -
    14 & L11-NGC-1569	      &  04:30:49.05 & +64:50:52.6 &-0.000347 &    2.89   &  10.69  (R) &     x     &		\\				%    67.7044110   64.847944    -
    15 & L11-NGC-1637	      &  04:41:28.17 & -02:51:28.6 & 0.002392 &    10.7   &  11.01  (R) &     x     &		\\				%    70.3674090   -2.857962    -
    16 & L06-NGC-1961	      &  05:42:04.64 & +69:22:42.3 & 0.013122 &    55.6   &  10.99  (V) &     x     &		\\				%    85.5193650   69.378437    HF
    17 & L13-UGC-3343	      &  05:45:24.64 & +72:21:22.4 & 0.003636 &    18.8   &  11.96  (I) &	    &		\\				%    86.3526670   72.356222    -
    18 & L12-NGC-2146	      &  06:18:37.71 & +78:21:25.3 & 0.002979 &    22.4   &   9.83  (R) &     x     &		\\				%    94.6571250   78.357028    -
    19 & L07-NGC-2273	      &  06:50:08.65 & +60:50:44.9 & 0.006138 &    30.9   &  10.52  (R) &     x     &		\\				%    102.536073   60.845806    -
    20 & L13-UGC-3691	      &  07:08:01.28 & +15:10:42.3 & 0.007348 &    36.5   &  12.60  (B) &	    &		\\				%    107.005333   15.178417    -
 \hline 																			
 \noalign{\smallskip}																		
   \noalign{\smallskip} 																	
   \hline																			
   \label{data1}
   \end{longtable}
   \normalsize
%   \end{onecolumn}

% \clearpage
% \begin{onecolumn}
 \scriptsize
 \tiny
 \begin{longtable}{clccclccccc}
 \caption{{\bf Observed line-parameters for a representative sample of 20 galaxies. The full set ot 376 measurements is available in electronic form
  at the CDS}}\\
 \hline
 \noalign{\smallskip}
    ID&   Obj		   & N   &  Texp &  ap   &  PA  			       &   EW[NII]1 &EWH$\alpha$ &   EW[NII]2 &  rms   &   class	    \\
      & 		   &	 &   s   &  "	 & deg  			       &  $\AA$   & $\AA$    &$\AA$	  &  Cnts  &			     \\
   (1)&    (2)  	   & (3) &  (4)  & (5)   & (6)  			       &      (7) &	 (8) &        (9) & (10)    &	(11)		    \\ 
 \hline
 \endfirsthead
 \hline
 \noalign{\smallskip}
    ID&   Obj		   & N   &  Texp &  ap   &  PA  			       &   EW[NII]1 &EWH$\alpha$ &   EW[NII]2 &  rms   &   class	    \\
     &  		   &	 &   s   &  "	 & deg  			       &  $\AA$   & $\AA$    &$\AA$	  &  Cnts  &			     \\   
   (1)&    (2)  	   & (3) &  (4)  & (5)   & (6)  			       &      (7) &	 (8) &        (9) & (10)    &	(11)		    \\ 
 \hline
 \endhead																    
 % &  L13-NGC-0205	& 3   &  300  &  2.5  &  90				   &   -      &  2.161   &     -      &  0.02  &   PAS  		\\
 % &  L13-M-32         & 3   &  200  &  2.5  &  90				   &   -      &  1.705   &     -      &  0.03  &   PAS  		\\
 % &  L13-M-31         & 3   &  300  &  2.5  &  90				   &   -      &  1.101   &     -      &  0.01  &   PAS  		\\
    1 &  L11-M33-HII	   & 3   &  180  &  2.0  &  90  			       &   -35.86 & -1314.0  &     -125.6 &  0.09  &   HII		    \\
    2 &  L05-CGCG-522-004  & 1   &  600  &  2.0  &  30  			       &  -0.53   &  -0.49   &    -2.92   &  0.04  &   wAGN		    \\
    3 &  L05-CGCG-522-093  & 1   &  600  &  2.0  & 120  			       &  -2.42   &  -9.18   &    -5.849  &  0.19  &   sAGN		    \\
    4 &  L05-CGCG-522-104  & 3   &  900  &  2.0  &  90  			       &  -2.32   &  -13.39  &    -8.115  &  0.11  &   sAGN		    \\
    5 &  L11-NGC-0784	   & 4   &  300  &  2.0  &   0  			       &   -	  &   -86.61 &     -	  &  0.25  &   HII		    \\
    6 &  L05-CGCG-522-106  & 3   &  600  &  2.0  & 150  			       &  -0.69   &  -3.23   &    -3.052  &  0.04  &   wAGN		    \\
    7 &  L09-NGC-0891	   & 3   &  300  &  2.5  &  90  			       &  -3.94   &  -36.45  &    -15.47  &  0.35  &   HII		    \\
    8 &  L00-NGC-0925	   & 7   &  300  &  2.0  &  85 (09,13)  		       &  -2.73   &  -45.23  &    -11.88  &  0.14  &   HII		    \\
    9 &  L09-UGC-1935	   & 3   &  420  &  2.0  &  74  			       &  -17.60  &  -21.6   &    -17.0   &  0.04  &   SEY1		    \\
   10 &  L00-NGC-1068	   & 4   &  470  &  2.0  &  90 (10,11)  		       &  -67.30  &  -111.2  &     -209.1 &  0.01  &   sAGN		    \\
   11 &  L00-NGC-1275	   & 6   &  300  &  2.0  &  90 (12,13)  		       &  -43.25  &  -71.56  &     -78.51 &  0.04  &   sAGN		    \\
   12 &  L09-UGC-2855	   & 3   &  300  &  2.0  & 105  			       &  -2.69   &  -23.55  &    -11.48  &  0.12  &   HII		    \\
   13 &  L11-NGC-1507	   & 4   &  180  &  2.0  &  10  			       &   -3.60  &   -79.16 &     -8.82  &  0.13  &   HII		    \\
   14 &  L11-NGC-1569	   & 2   &  180  &  2.0  &  90  			       &   -	  &   -116.1 &     -3.28  &  0.03  &   HII		    \\
   15 &  L11-NGC-1637	   & 2   &  300  &  2.0  &  10  			       &   -7.53  &   -37.21 &     -23.05 &  0.03  &   sAGN		    \\
   16 &  L06-NGC-1961	   & 3   &  300  &  2.0  &  90  			       &  -6.05   &  -10.01  &    -19.20  &  0.01  &   sAGN		    \\
   17 &  L13-UGC-3343	   & 3   &  360  &  2.5  &  79  			       &  -3.74   &  -71.89  &    -9.31   &  0.17  &   HII		    \\
   18 &  L12-NGC-2146	   & 3   &  300  &  2.0  &  90  			       &  -7.94   &  -53.66  &    -24.78  &  0.04  &   HII		    \\
   19 &  L07-NGC-2273	   & 1   &  180  &  2.0  &  90  			       &  -8.89   &  -29.57  &    -27.93  &  0.04  &   sAGN		    \\
   20 &  L13-UGC-3691	   & 3   &  600  &  2.5  &  63  			       &   -	  &  -8.71   &    -4.38   &  0.23  &   HII		    \\
   \hline
   \noalign{\smallskip}
   \noalign{\smallskip}
   \hline
   \label{data2}
   \end{longtable}
   \normalsize
   \end{onecolumn}

      \clearpage
       \begin{figure*}
      \begin{center}
      \includegraphics[width=6cm]{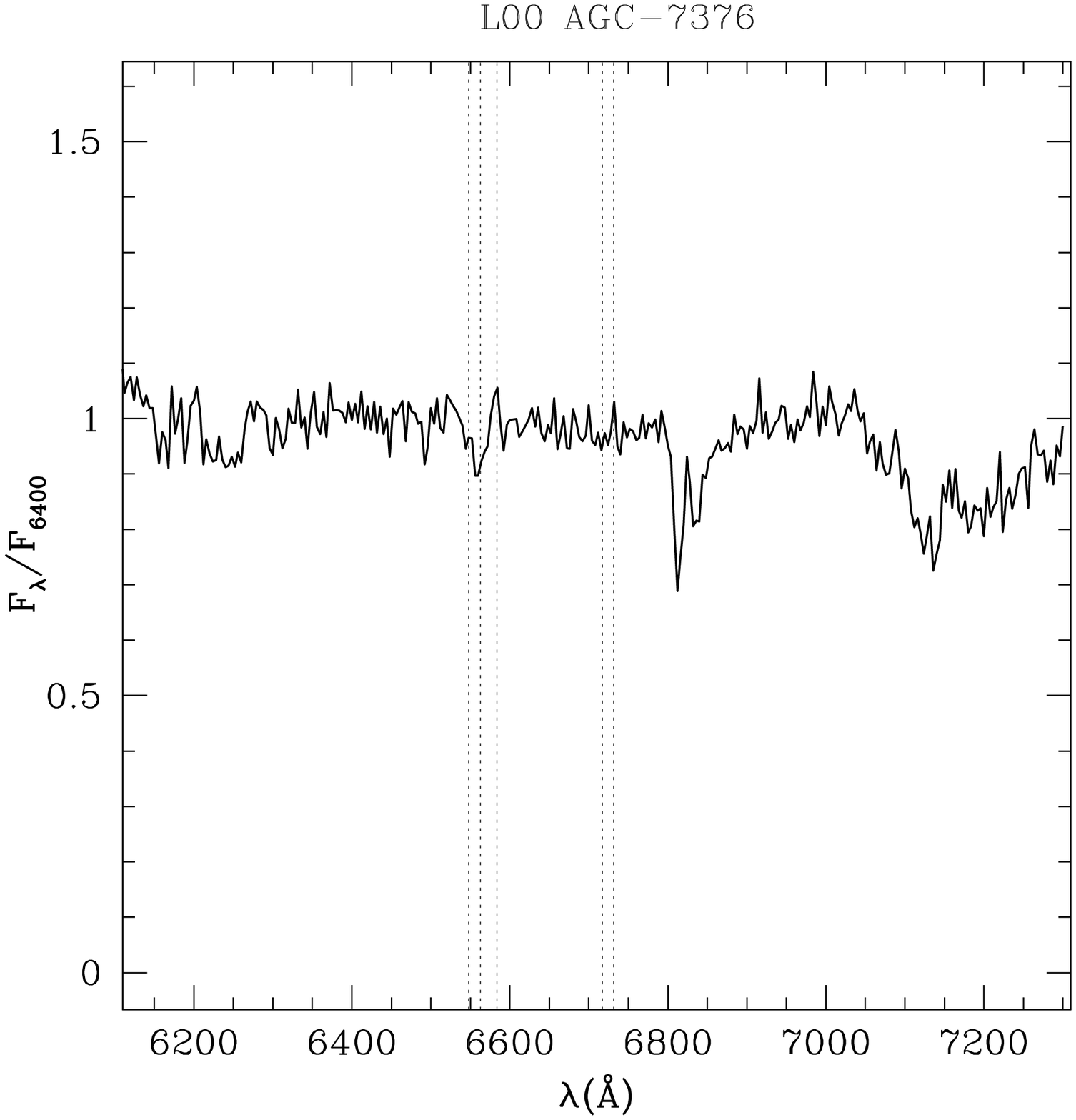}
      \includegraphics[width=6cm]{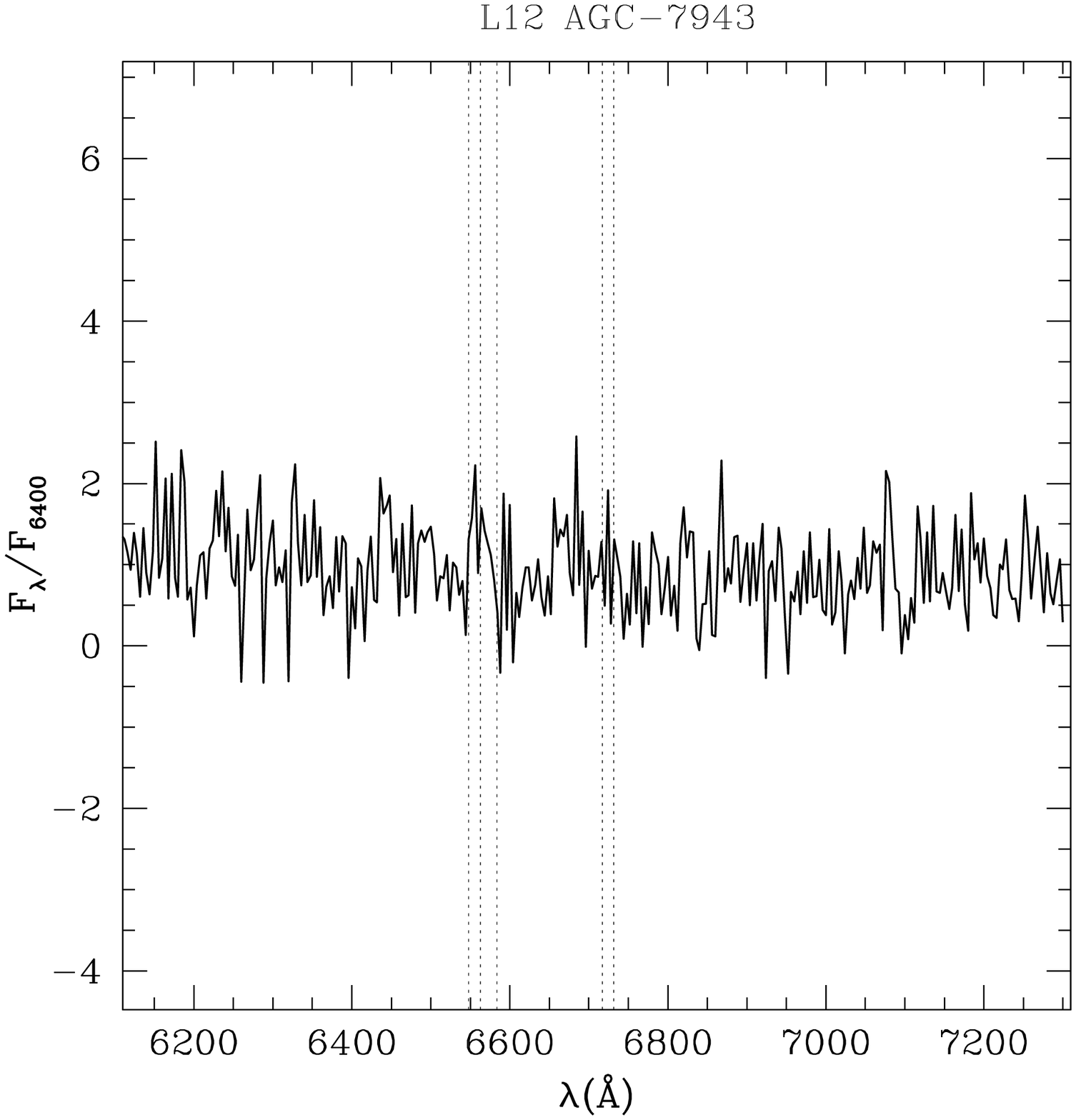}
      \includegraphics[width=6cm]{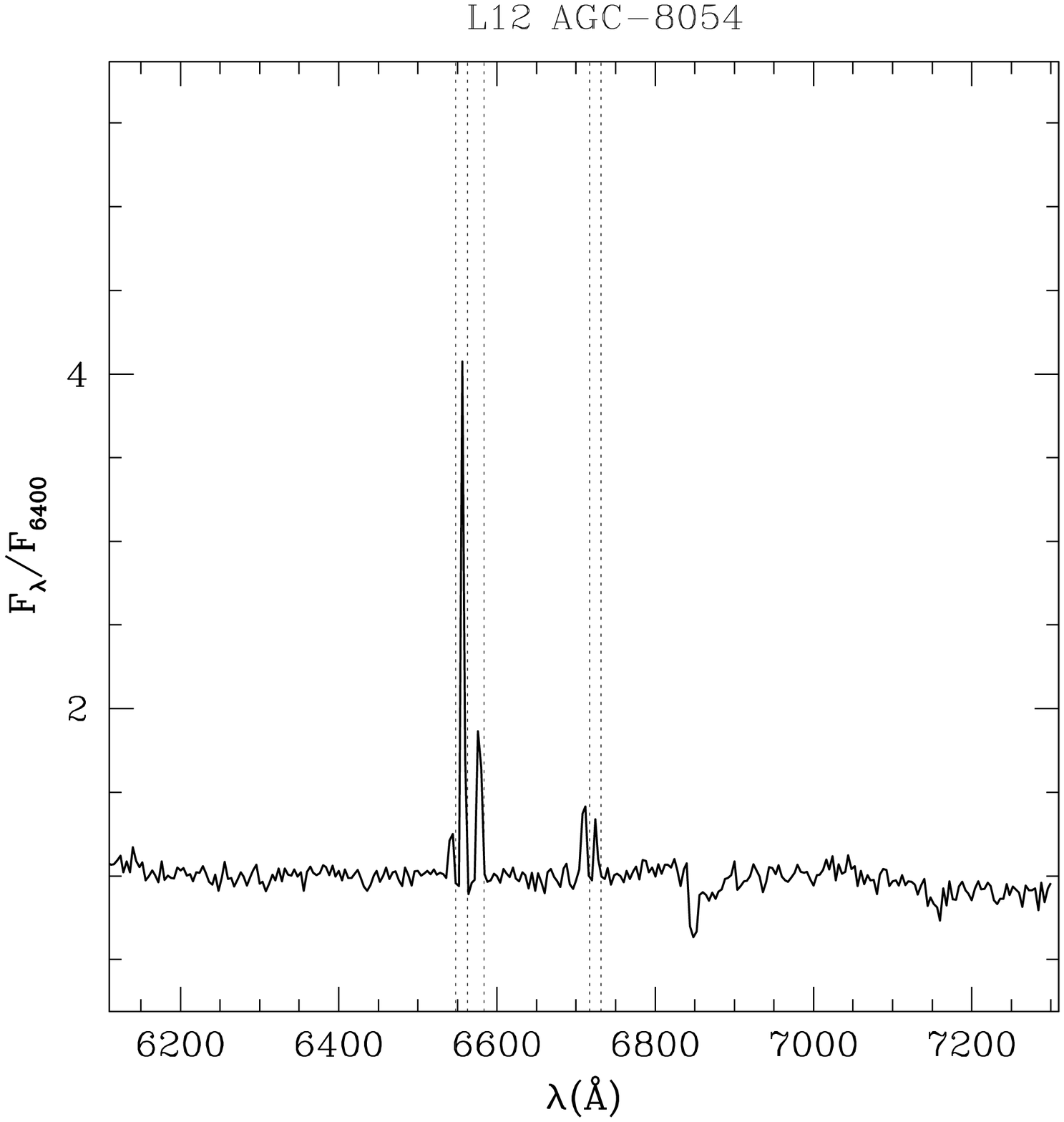}\\
      \includegraphics[width=6cm]{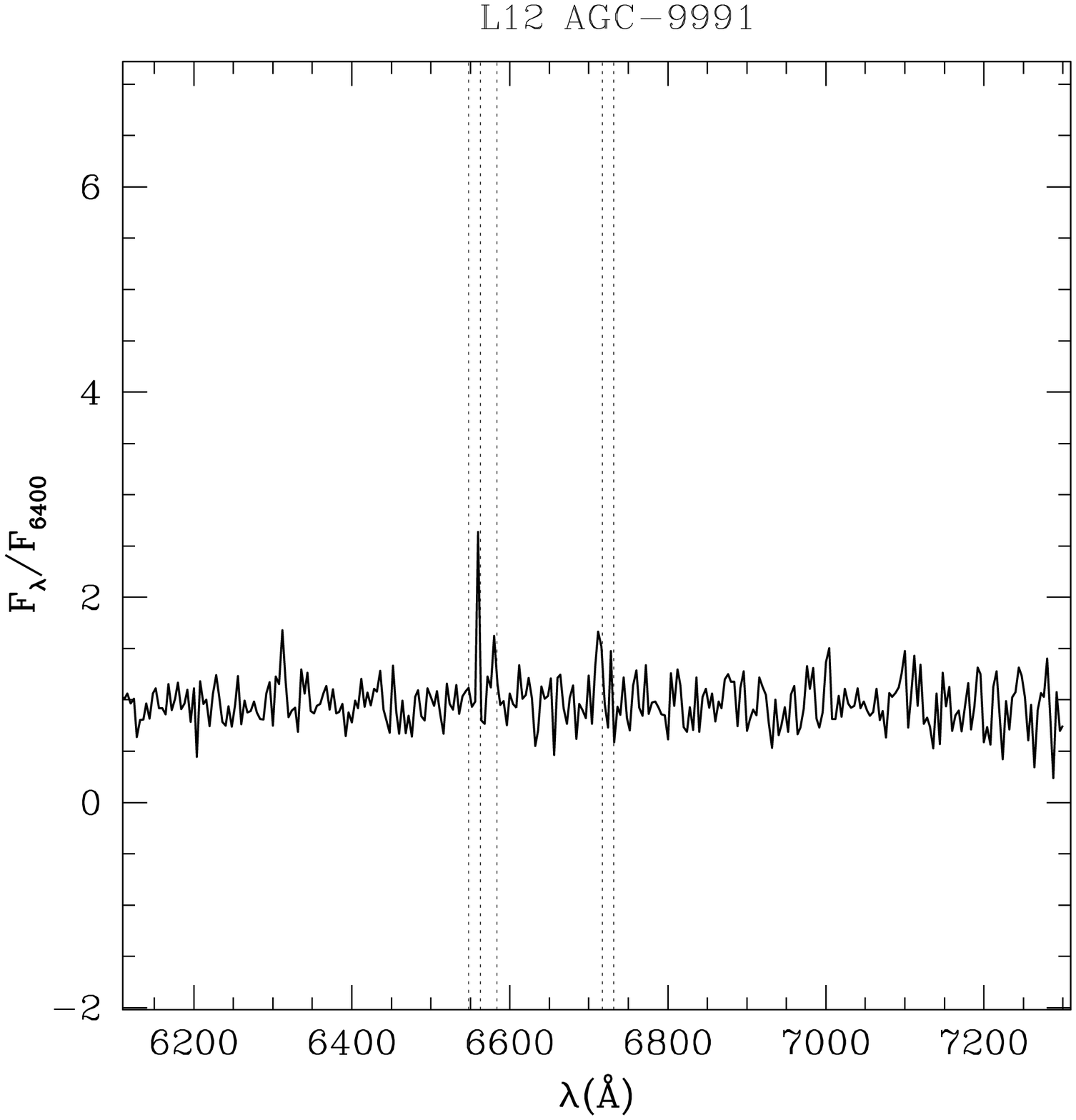}
      \includegraphics[width=6cm]{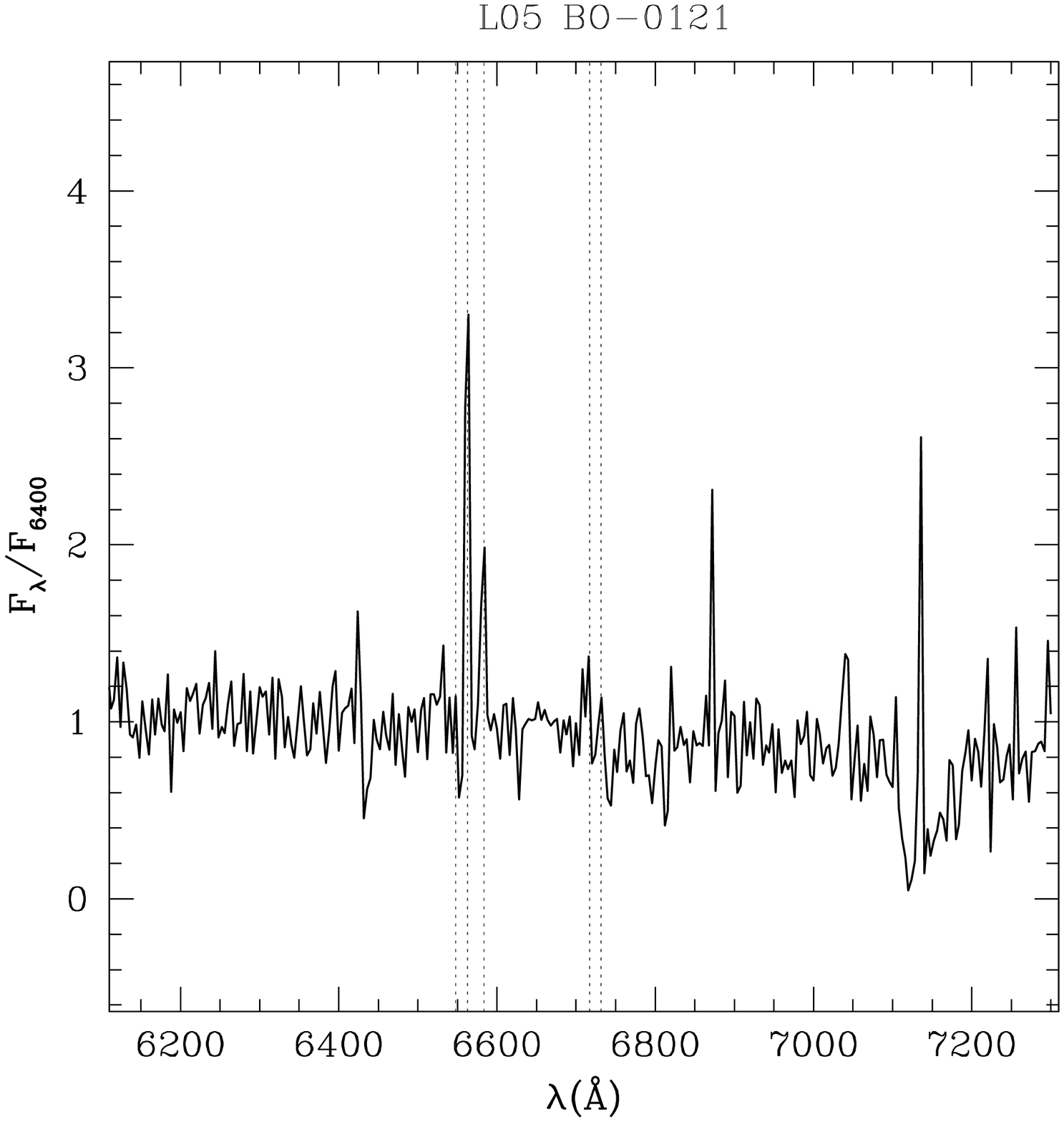}
      \includegraphics[width=6cm]{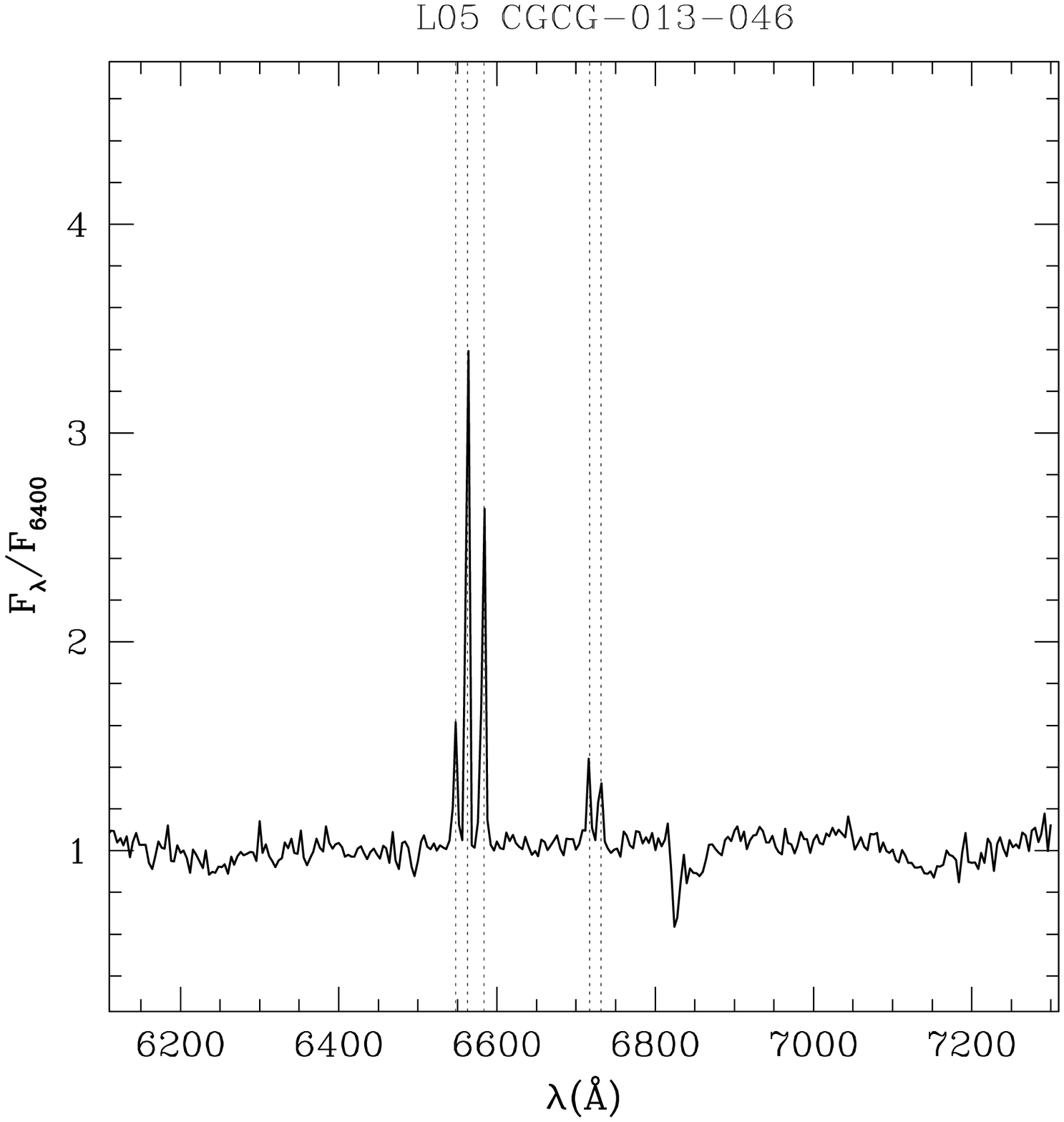}\\
      \includegraphics[width=6cm]{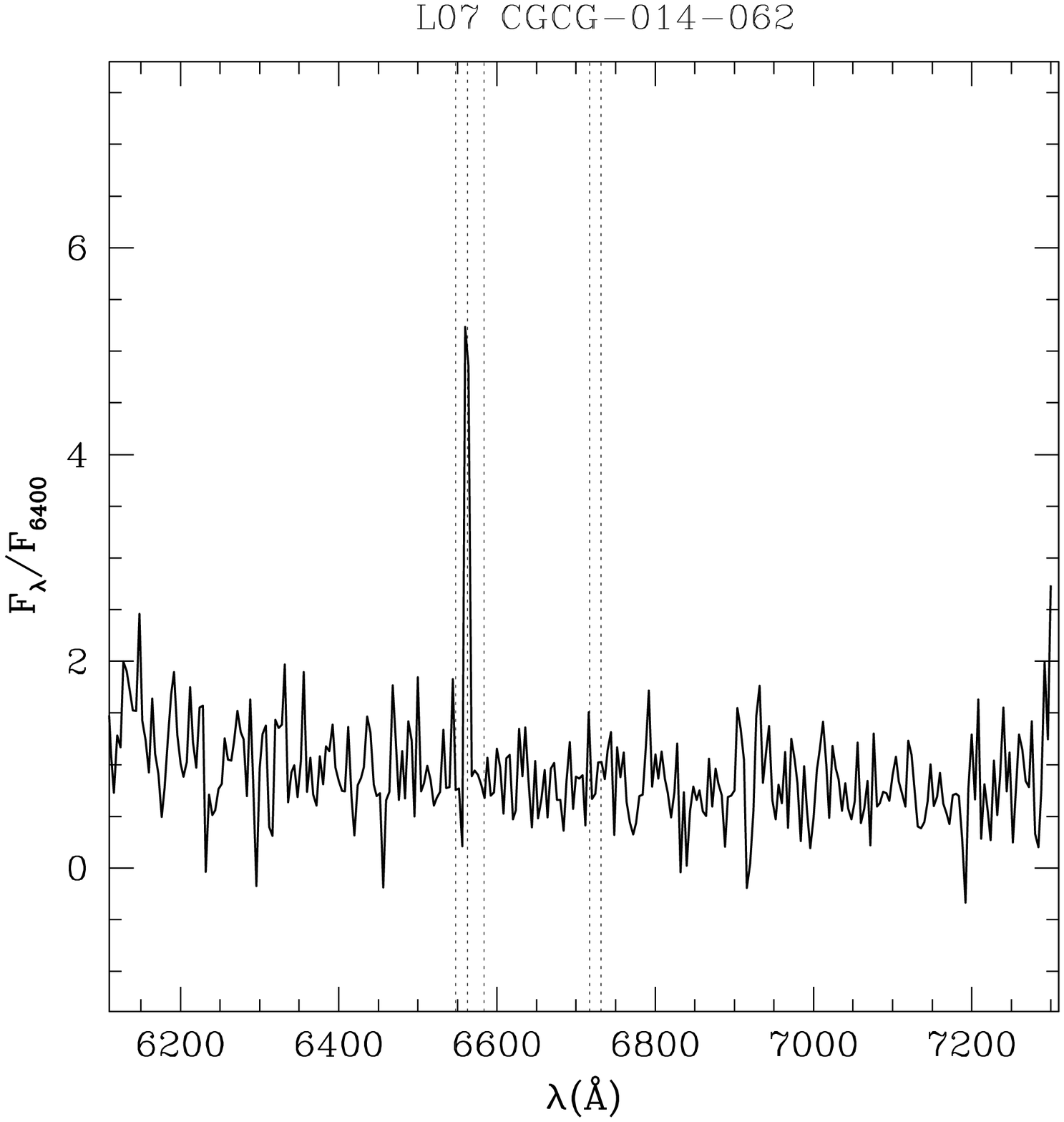}
      \includegraphics[width=6cm]{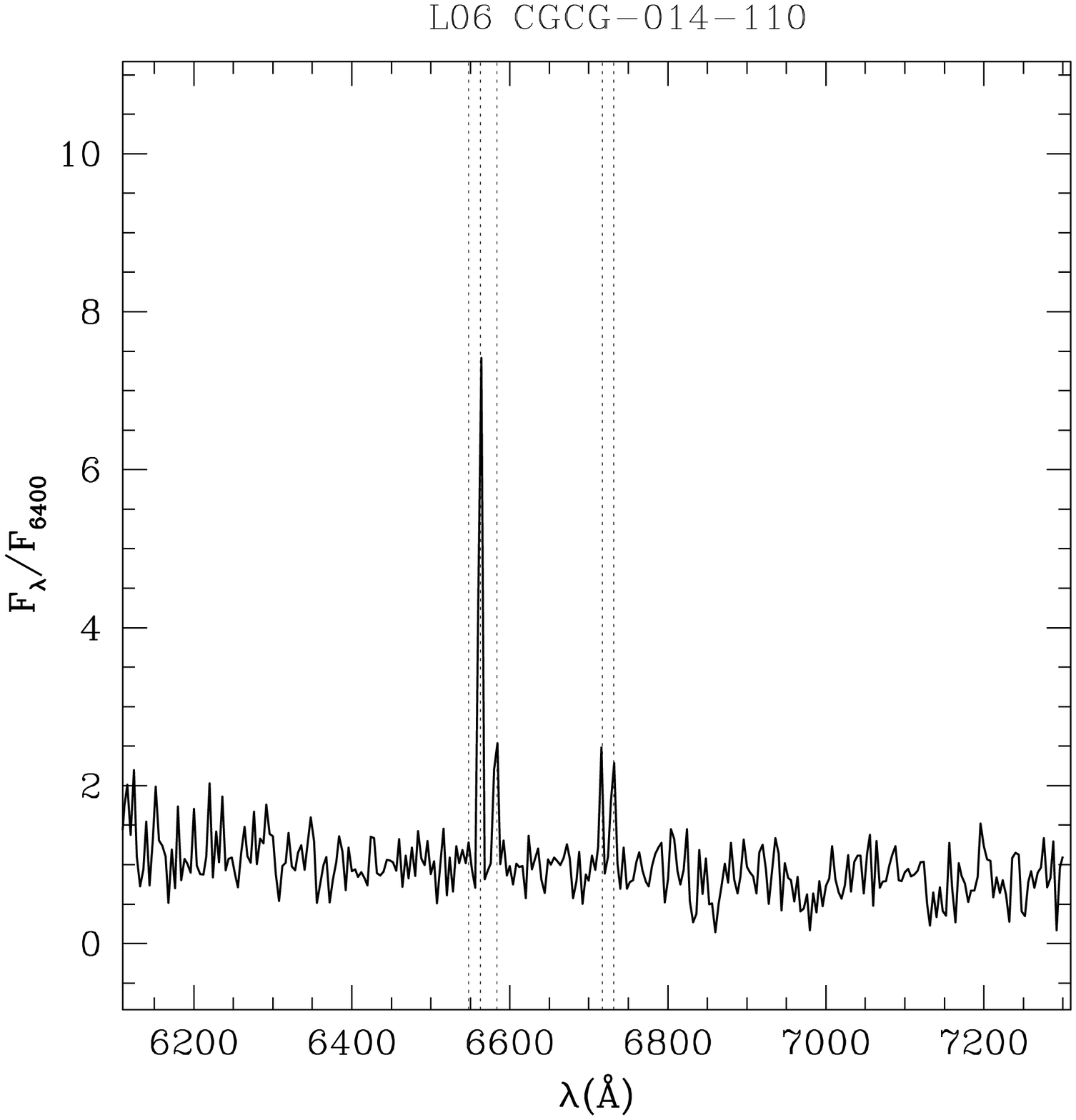}
      \includegraphics[width=6cm]{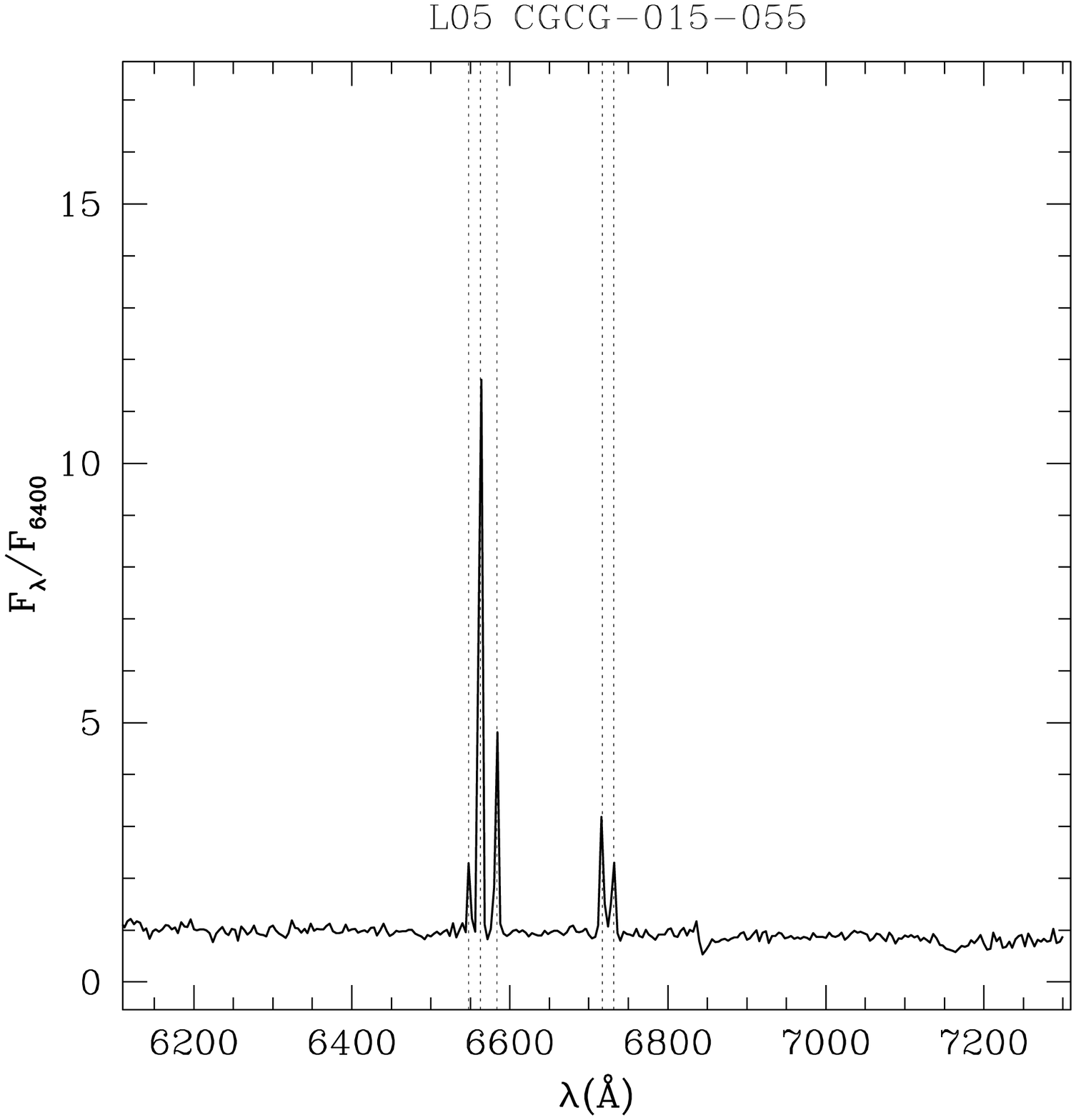}\\
      \includegraphics[width=6cm]{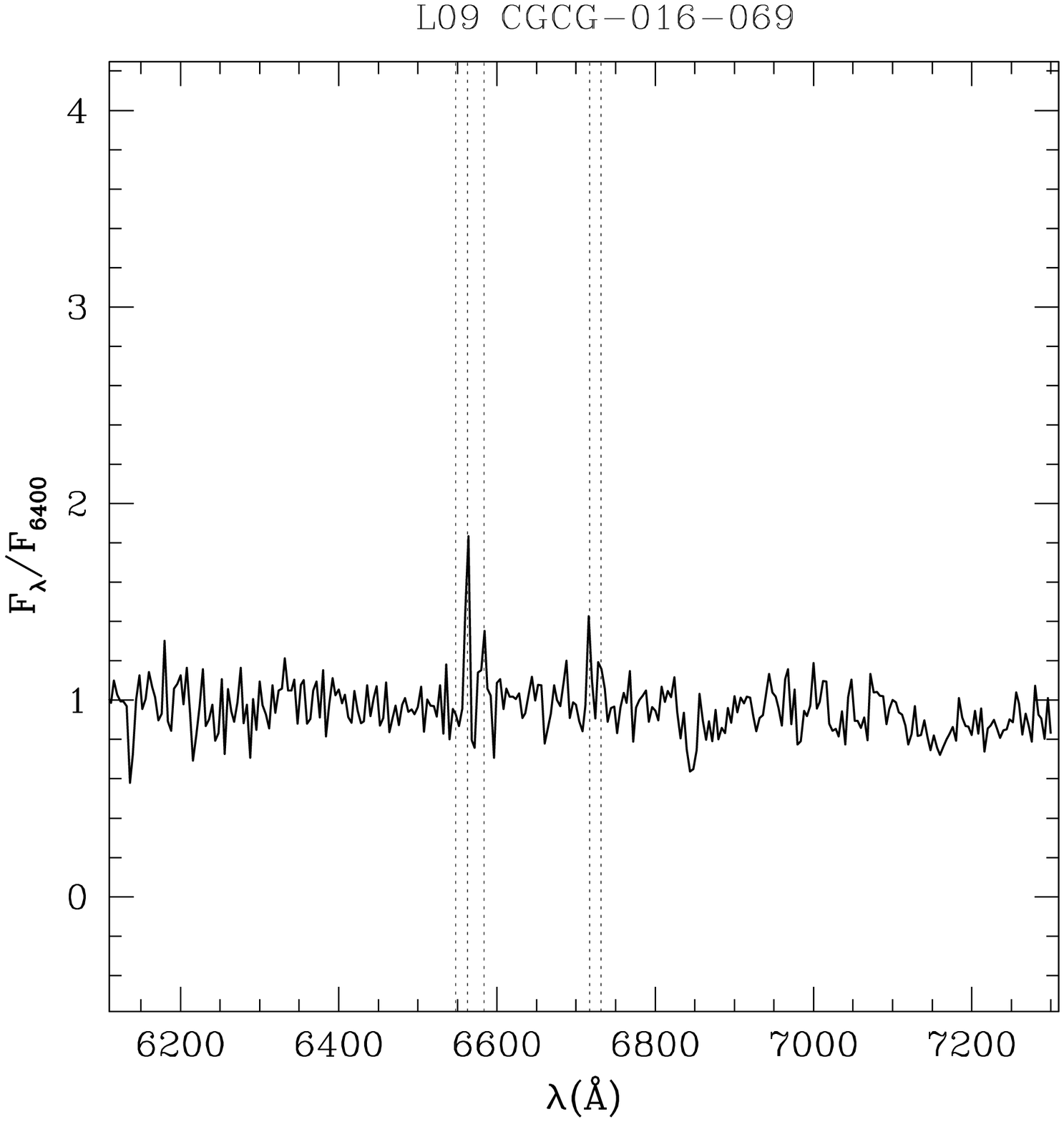}
      \includegraphics[width=6cm]{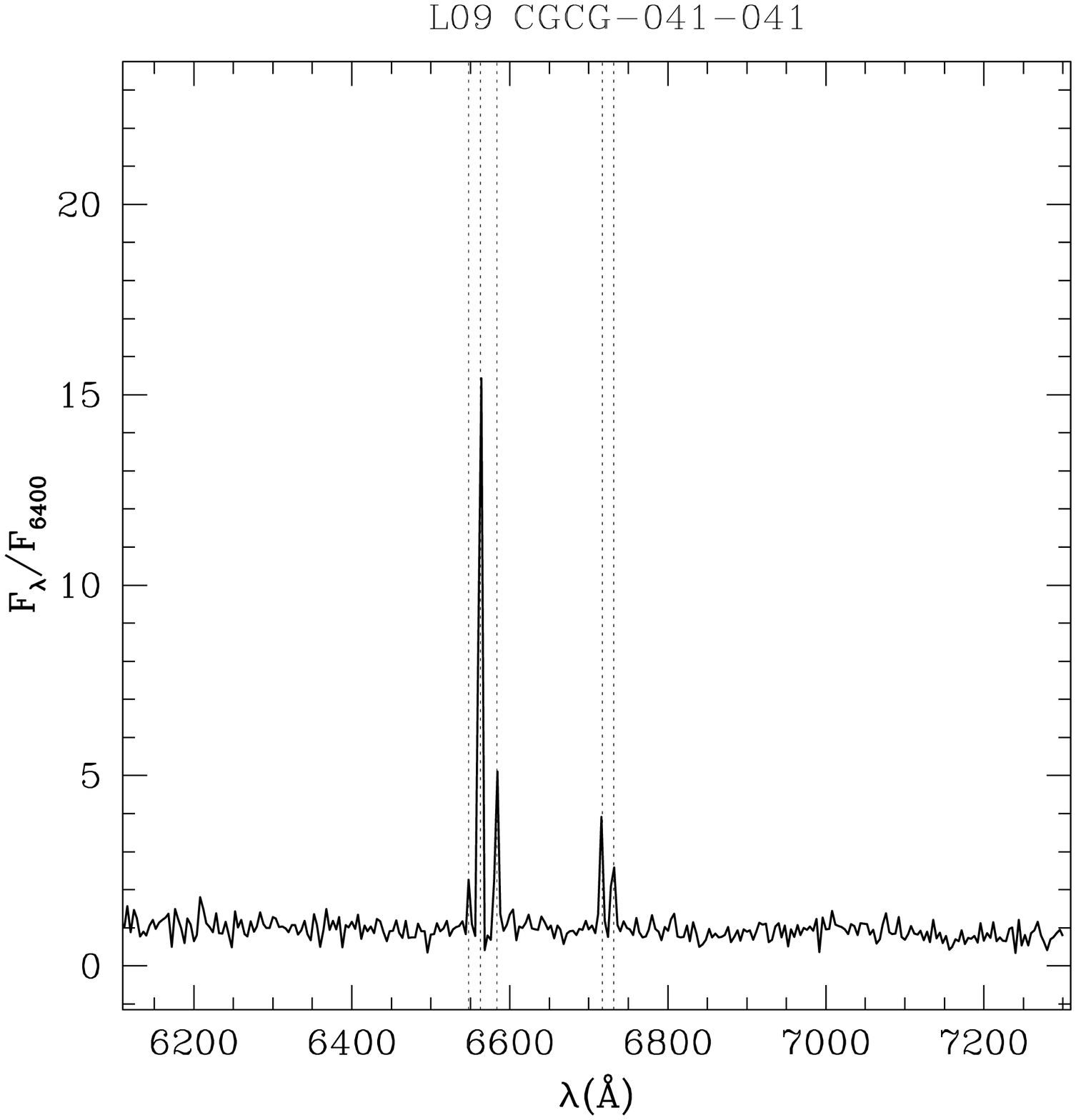}
      \includegraphics[width=6cm]{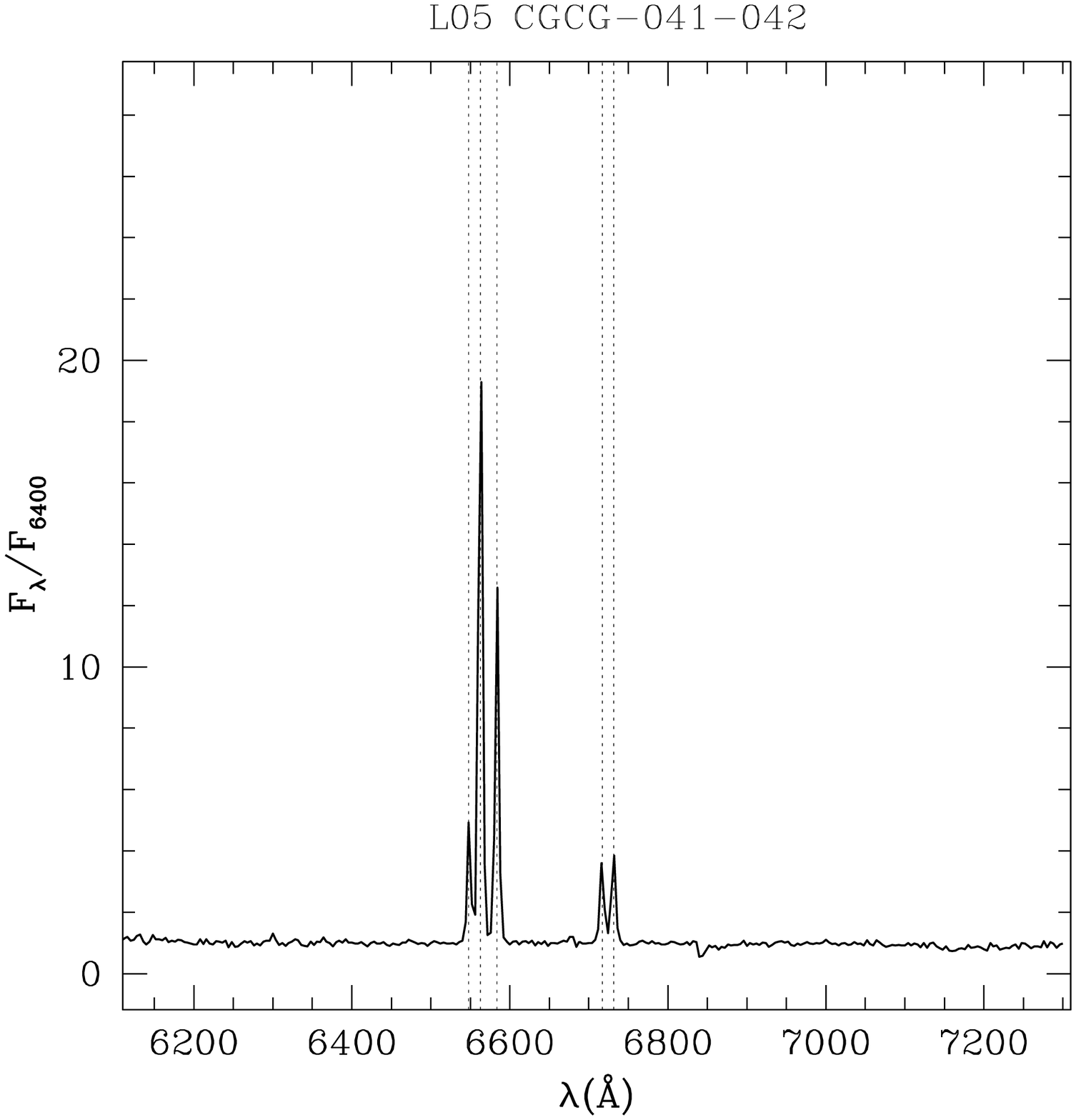}\\
    \caption{{\bf Red restframe-spectra     
    (from 6100 $\AA$ to 7300 $\AA$) obtained with the red-channel grism in this work, normalized to the continuum near
    6400 \AA. The vertical dashed lines give the position of $[NII]_{6548}$, H$\alpha$ ($\lambda$ 6562.8), 
    $[NII]_{6583}$, $[SII]_{6717}$, $[SII]_{6731}$.  
    The first three digits of the label give the year of observation. 
    L00 marks those spectra that are a median of two or more spectra obtained in different runs.
    Galaxies are listed in order of catalog names. A representative sample of 12 spectra is shown. 
    The full set ot 376 spectra is available in electronic form at the CDS.}}
    \label{spectra}
    \end{center}
    \end{figure*}

       \begin{figure*}
       \begin{center}
       \includegraphics[width=6cm]{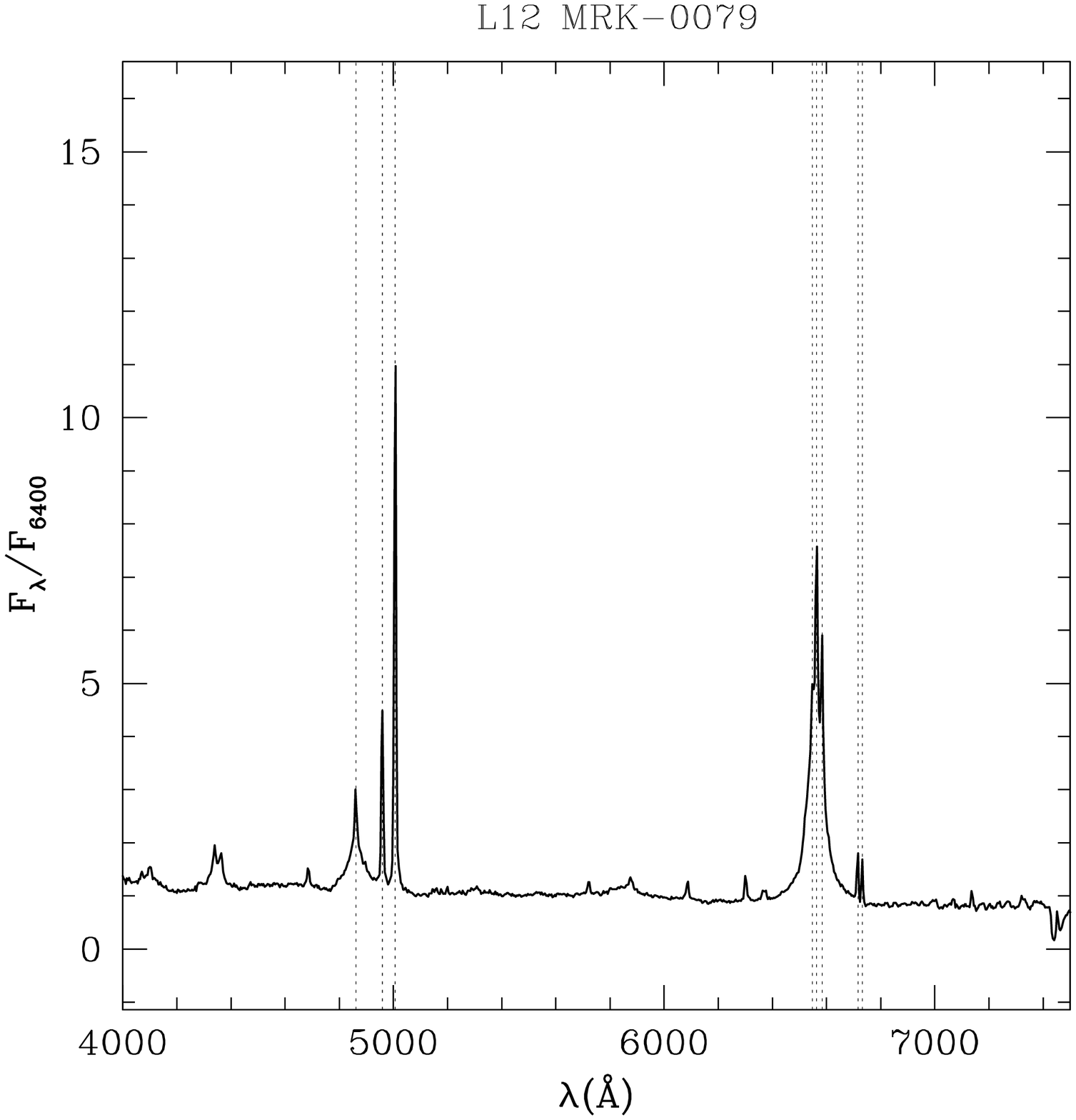} 
       \includegraphics[width=6cm]{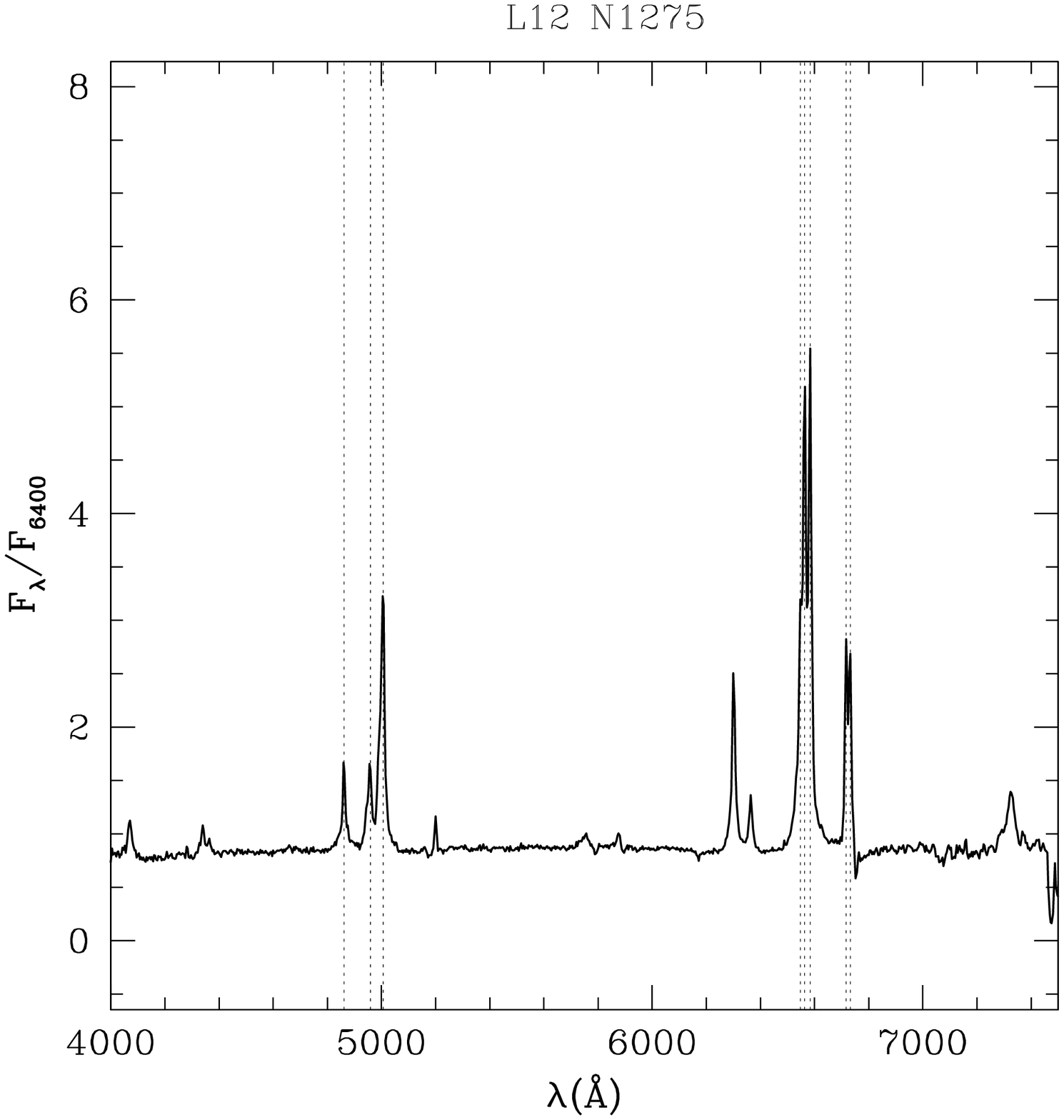}  
       \includegraphics[width=6cm]{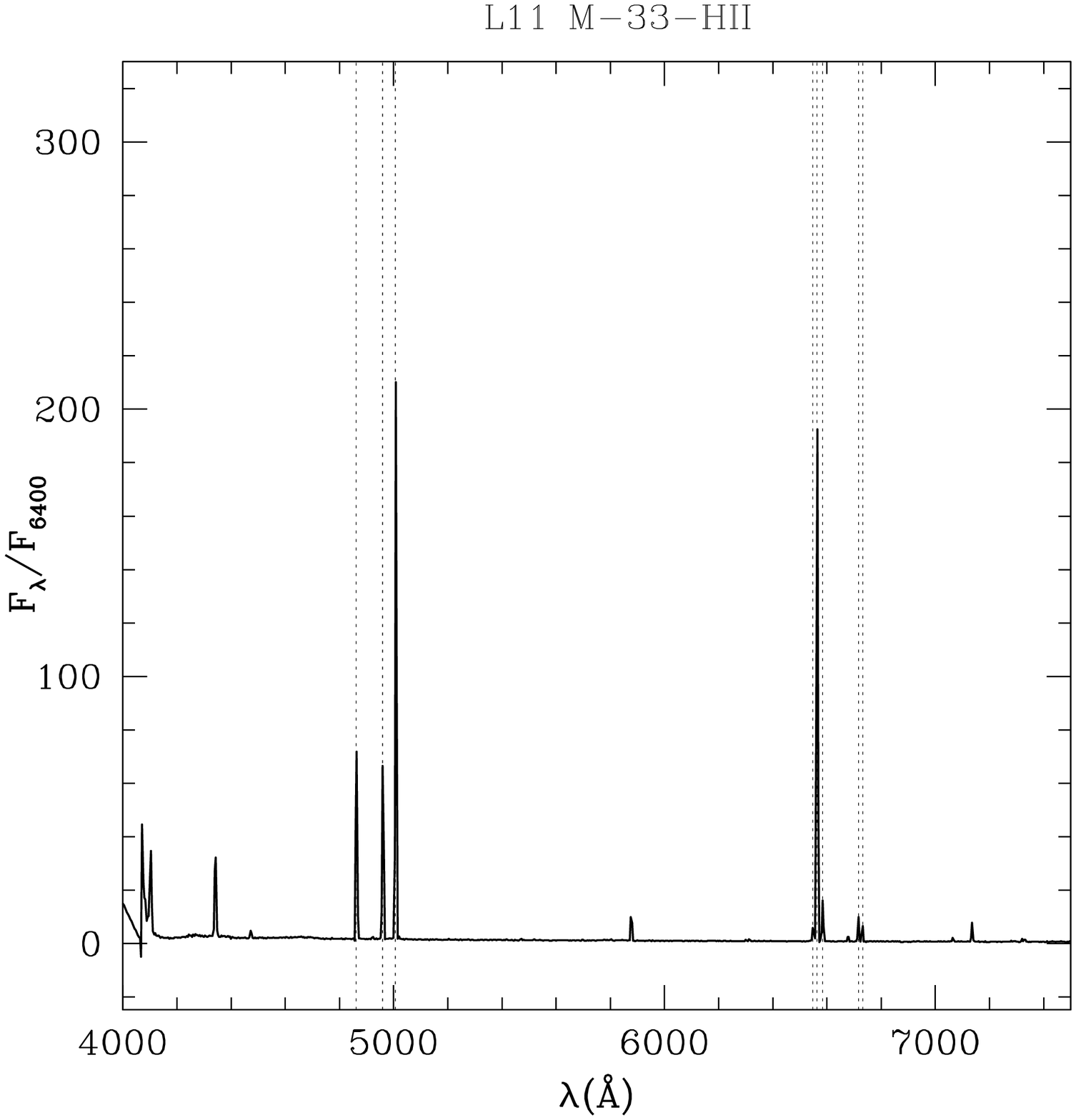}
       \caption{Spectra obtained with the blue- and red-channel grisms normalized to the continuum near 6400 \AA.}
    \label{G7}
    \end{center}
    \end{figure*}

       \begin{figure*}
       \begin{center}
      \includegraphics[width=6cm]{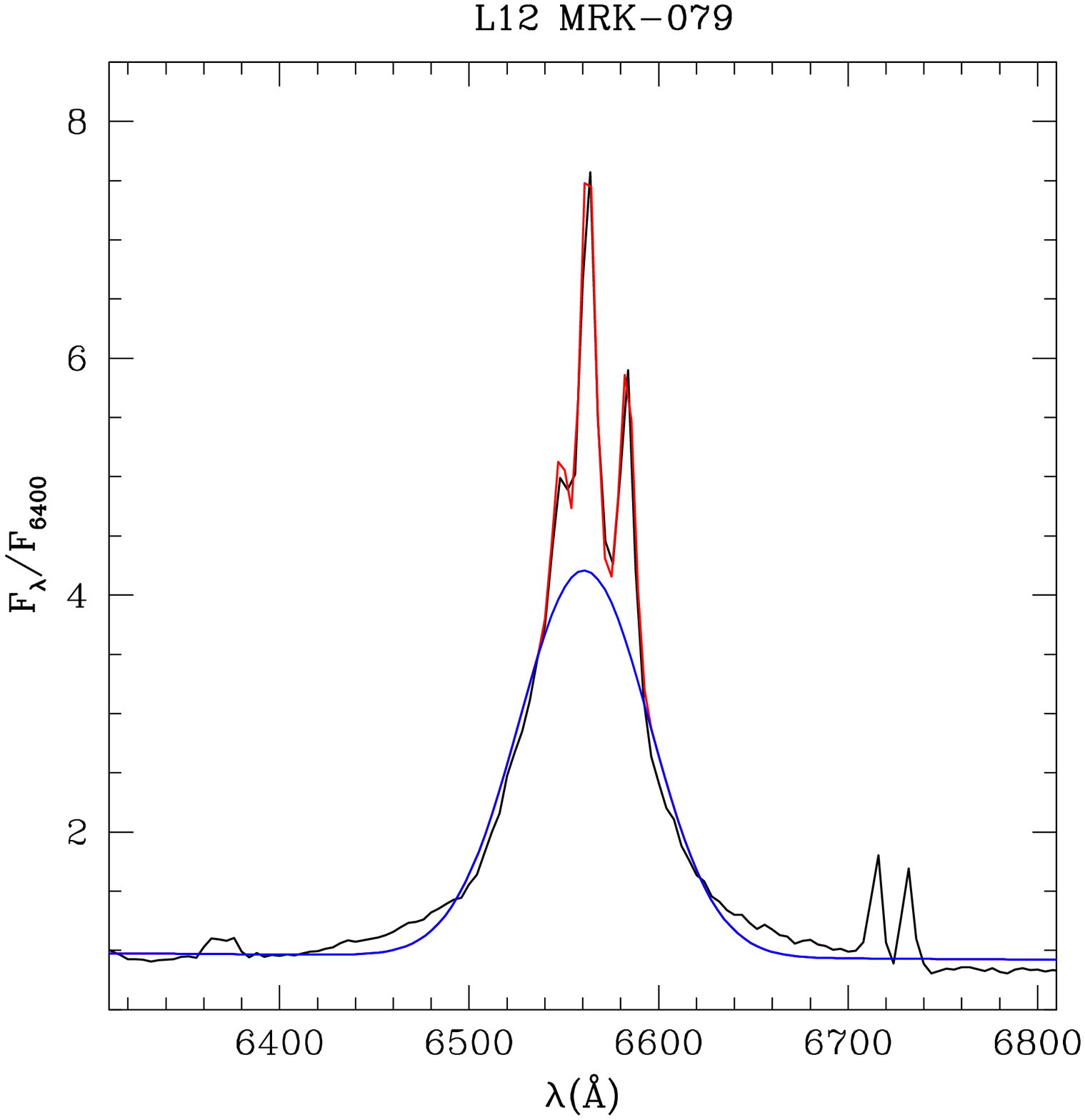} 
      \includegraphics[width=6cm]{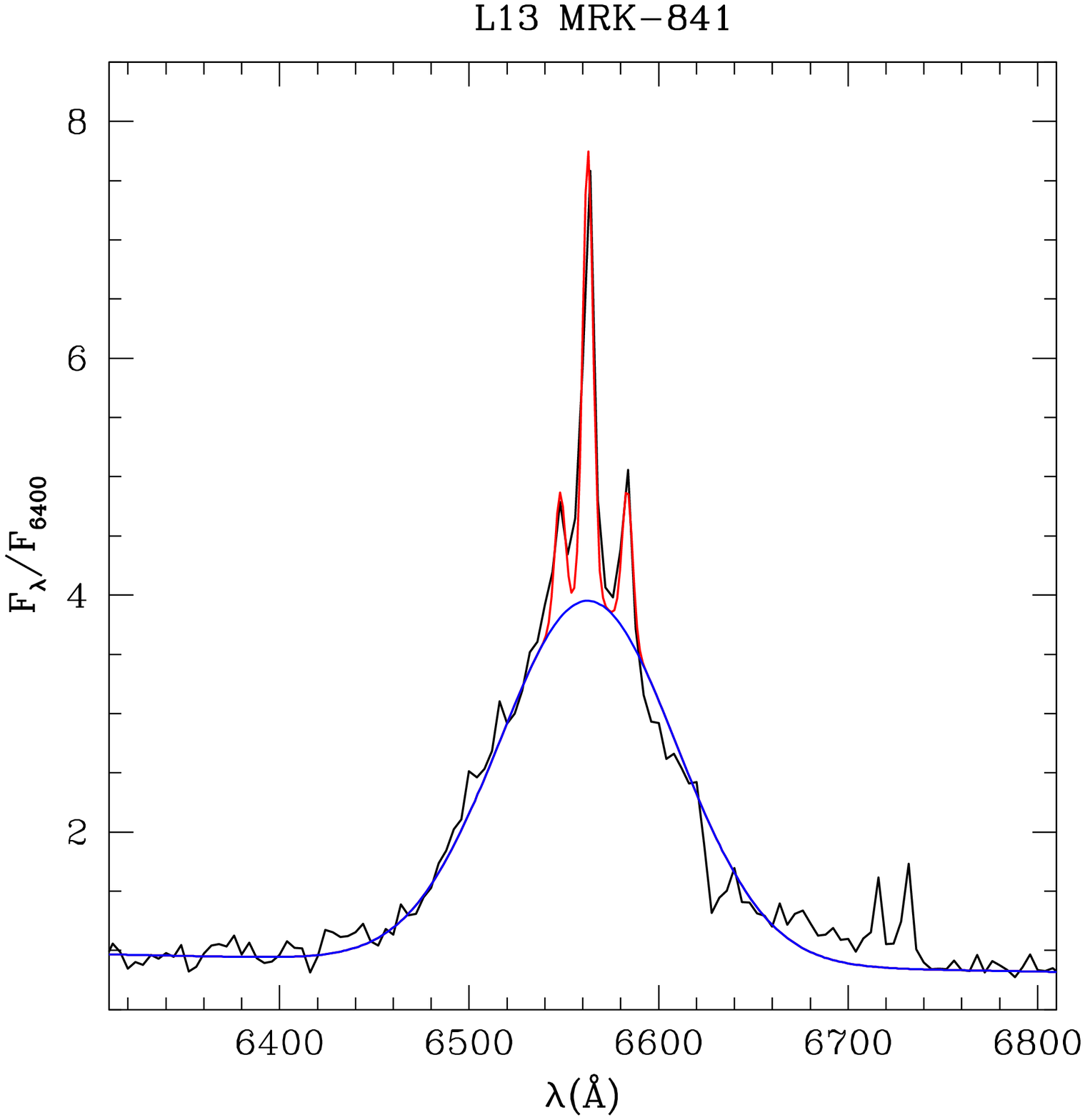}\\ 
      \includegraphics[width=6cm]{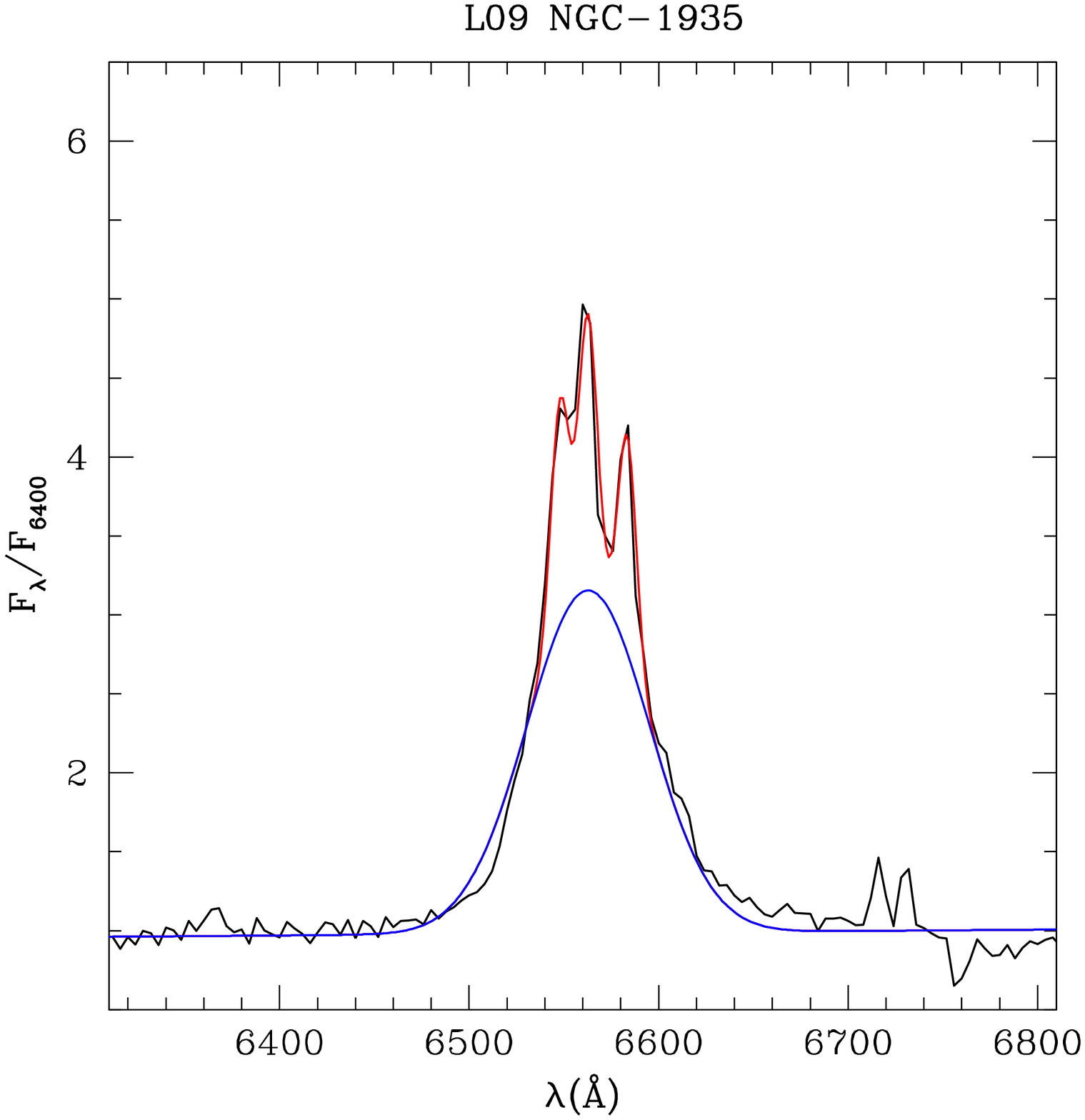}
      \includegraphics[width=6cm]{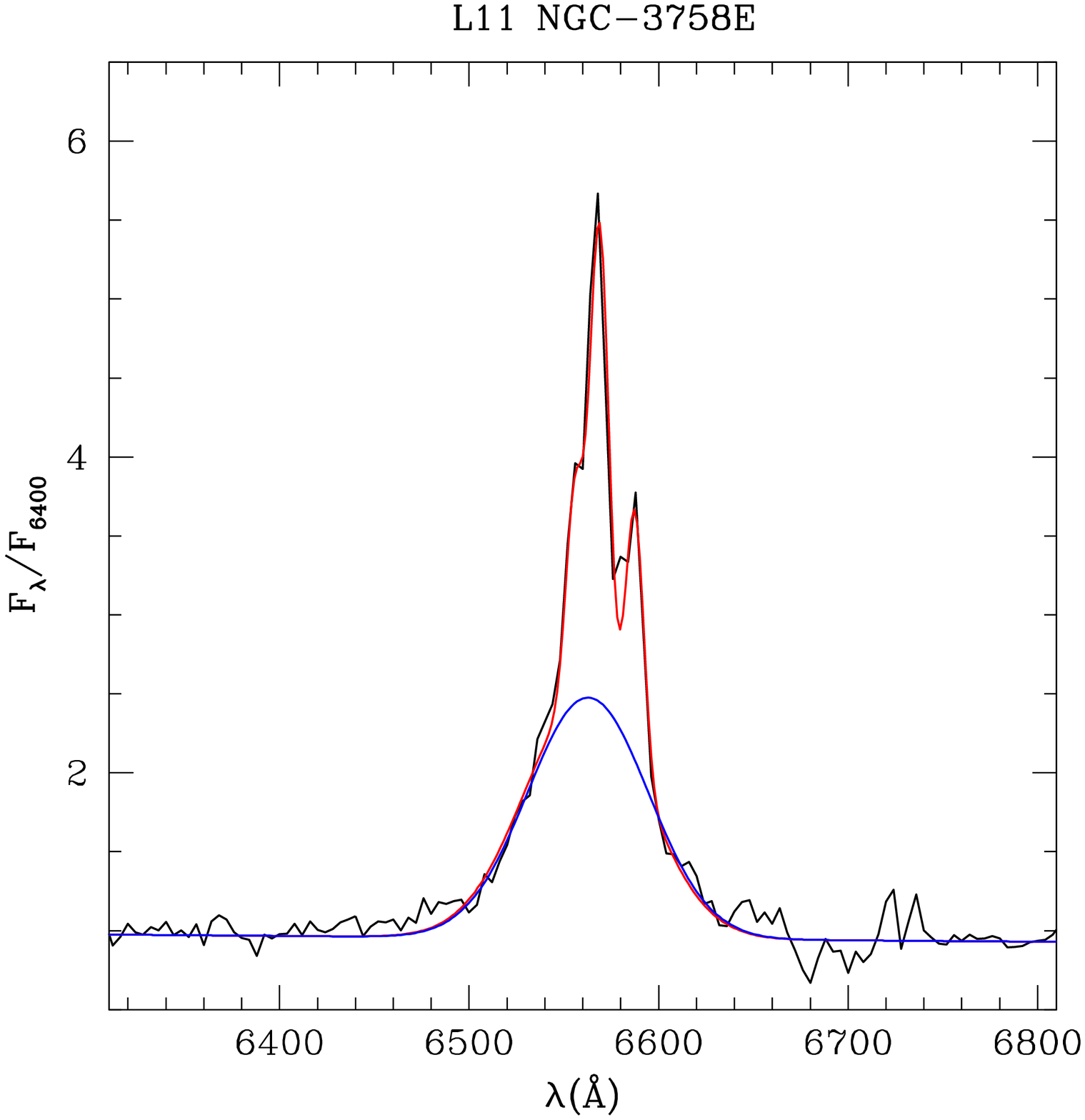}\\
       \caption{Spectra of the four Seyfert 1 galaxies in the sample. The
	 black lines refer to the original spectra. The red lines show the
	 full fit (continuum + broad+narrow lines), while the blue lines
	 highlight the contribution of the broad H$\alpha$ component.}
    \label{broad}
    \end{center}
    \end{figure*}
\end{document}